\documentclass{article}
\usepackage{amsmath}
\usepackage{amssymb}
\usepackage{amsthm}
\usepackage[english,russian]{babel}
\usepackage[final]{graphicx}

\def\leq{\leqslant}

\def\geq{\geqslant}
\def\RR{\mbox{\rm I\hspace*{-.2em}R}}
\begin{document}

\begin{center}
\large\sc Minimal Committee Problem\\ for Inconsistent Systems of\\ Linear Inequalities on the Plane
\end{center}

\begin{center}
K. S. Kobylkin

Institute of Mathematics and Mechanics,\\ Ural Branch, Russian Academy of Sciences,

ul. S. Kovalevskoi 16, Ekaterinburg, 620219 Russia

e-mail: \verb'kobylkinks@gmail.com'
\end{center}

{\bf Abstract.}

A representation of an arbitrary system of strict linear inequalities in $\RR^n$ as a system of points is
proposed. The representation is obtained by using a so-called polarity. Based on this representation an algorithm
for constructing a committee solution of an inconsistent plane system of linear inequalities is given. A solution
of two problems on minimal committee of a plane system is proposed. The obtained solutions to these problems can
be found by means of the proposed algorithm.

DOI: 10.1134/S1054661806040201

\section{INTRODUCTION}

A problem of two finite sets separation with several (possibly small number of) hyperplanes is considered. There
are different ways in which two subsets in $\RR^n$ can be separated. We focus on the one that involves majority
voting principle. Let ${\cal{A}}$ and ${\cal{B}}$ be arbitrary subsets of $\RR^n,$ where $n$ is an arbitrary
positive integer.

{\bf Definition 1.} A {\it committee of linear functions} \cite{mazurov}, which separates two subsets of points
${\cal{A}}$ and ${\cal{B}},$ is a finite collection (with possible repetitions) of linear functions such that at
any point of ${\cal{A}}$ (res\-pec\-ti\-vely of ${\cal{B}}$) more than half of functions of this collection are
positive (respectively negative) counting multiplicity.

It generalizes separating hyperplane notion in the case where the subsets are inseparable, i.e. if they are
linearly separable by a hyperplane $(w,x)=\alpha$ the set $\{f(\cdot)=(w,\cdot)-\alpha\}$ of the one function is a
committee.

Separating committee notion has clear geometrical interpretation (Fig.~1). Two linearly inseparable sets
${\cal{A}}$ and ${\cal{B}}$ are shown by three points and three crosses respectively. There is a committee of
three functions separating them which is shown in the figure by three straight lines. For each line we have its
positive and negative sides defined by corresponding half-planes. Every point of ${\cal{A}}$ (respectively, of
${\cal{B}})$ is contained in intersection of two positive (respectively, negative) half-planes among the three.

\begin{figure}[htb]\label{ksfig1}
\begin{center}
\includegraphics[width=6cm]{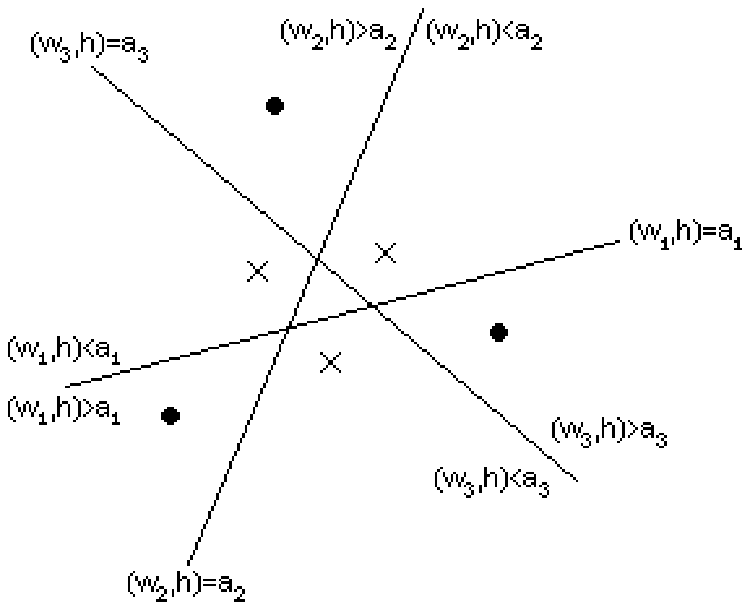}
\caption{}
\end{center}
\end{figure}

Finding a committee of $q_0$ functions that separates ${\cal{A}}$ and ${\cal{B}}$ given positive integer $q_0$ is
equivalent to learning of a particular type of two layer perceptrons. Consider the perceptron (Fig.~2) whose
neurons are threshold units with a single neuron in output layer which sums up the outputs of $q_0$ neurons of one
hidden layer where training set is given by the subsets ${\cal{A}}$ and ${\cal{B}}.$ Here the $i$th hidden neuron
computes the threshold function $z_i(\cdot)={\rm{sgn}}((w_i,\cdot)-\alpha_i),\,\,i=1,\ldots,q_0,$ and the output
neuron realizes the threshold function $g(z)={\rm{sgn}}(\sum\limits_{i=1}^{q_0} z_i),$ where

\[{\rm{sgn}}\, x = \left\lbrace
           \begin{array}{c l}
              $--1,$ & \text{$x<0,$}\\
              $0,$ & \text{$x=0,$}\\
              $1,$ & \text{$x>0.$}
           \end{array}
         \right. \]
Input layer consists of $n$ nodes $x_1,\ldots,x_n\in\RR,$ and each of them is connected to every hidden neuron.
Then $\{f_i(\cdot)=(w_i,\cdot)-\alpha_i\}_{i=1}^{q_0}$ is a separating committee iff corresponding perceptron's
output is $+1$ at any point of ${\cal{A}},$ and $-1$ at any point of ${\cal{B}}.$ Thus, the problem is reduced to
training of the perceptron that implies adjusting its hidden neurons weights $\{w_i\}_{i=1}^q$ and thresholds
$\{\alpha_i\}_{i=1}^q.$

\begin{figure}[htb]\label{ksfig2}
\begin{center}
\includegraphics[width=6cm]{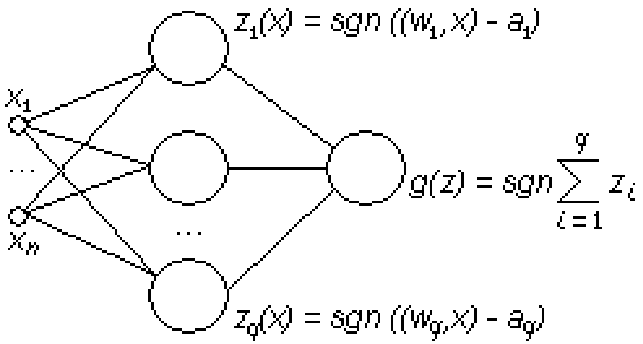}
\caption{}
\end{center}
\end{figure}

{\bf Definition 2.} A {\it committee} \cite{mazurov} of a system of strict linear inequalities
\begin{equation}\label{ks00}
(c_j,x)>b_j,\,\, j=1,\ldots,m,\,\, c_j,x\in\RR^n,\,\, b_j\in\RR
\end{equation}
is a finite collection (with possible repetitions) of vectors of $\RR^n$ such that each inequality of the system
is satisfied by more than half of members of the collection. A committee with the minimal number of elements
(taking into account their multi\-pli\-ci\-ty) for a given system $(\ref{ks00})$ is called a {\it minimal
committee}.

Every committee \cite{mazurov} that separates subsets ${\cal{A}}$ and ${\cal{B}}$ can be easily transformed to a
committee of a system
\begin{equation}
(c,z)>0, c\in {\cal{A}}'\cup (-{\cal{B}}'),\, z\in\RR^{n+1},
\end{equation}
where ${\cal{A}}'=\{[a,1]:a\in {\cal{A}}\}$ and ${\cal{B}}'=\{[b,1]:b\in {\cal{B}}\}.$ The statement extends to
the case of nonhomogeneous system $(\ref{ks00})$ as follows. Suppose that $b_j\neq 0,$ $j=1,\ldots,m.$ A
collection $\{x^i\}_{i=1}^q\subset\RR^n$ is a committee of $(\ref{ks00})$ iff the collection
$\{f_i(\cdot)=(x^i,\cdot)-1\}_{i=1}^q$ is a committee separating the subsets ${\cal{A}}=\left\{\frac{c_j}{b_j}:
b_j>0\right\}$ and ${\cal{B}}=\left\{\frac{c_j}{b_j}: b_j<0\right\}\cup\{0\}.$ Thus, finding a committee of the
system $(\ref{ks00})$ is reduced to constructing a collection of linear functions with positive constant terms
such that at any point of ${\cal{A}}$ (res\-pec\-ti\-vely of ${\cal{B}}$) more than half of functions of this
collection are positive (respectively negative).

We focus on the problem

{\bf Problem.} Find a minimal committee of the system $(\ref{ks00}).$

In the case of a system of strict homogeneous linear inequalities on the plane, this problem is completely solved
\cite{mazurov}. The case of nonhomogeneous plane systems, as well as the case of homogeneous systems for $n>2,$
remains unsolved. Moreover the problem becomes $NP$-hard for $n=3$ \cite{kh} even in the case of homogeneous
system if no additional conditions have been imposed on $(\ref{ks00}).$

In this paper, the representation of an arbitrary system of strict linear inequalities in $\RR^n$ is proposed as a
system of points being the union of two sets of points ${\cal{A}}$ and ${\cal{B}},$ where ${\cal{B}}$ contains a
zero point. We consider the problem of constructing a minimal committee of a system of strict linear inequalities
on the plane. In this paper, an algorithm for constructing a committee is proposed, which is based on the
representation of this system as a system of points. The solution of two problems on the minimal committee of such
systems is given, and the obtained solutions to these problems can be found by using of the proposed algorithm.
\clearpage

\section{REPRESENTATION OF AN ARBITRARY \\SYSTEM OF LINEAR INEQUALITIES \\ AS A SYSTEM OF POINTS}

\begin{center}
{\sc Definition of polarity}
\end{center}

Let $S$ be a unit sphere centered at $0$ that is defined in the $n$-dimensional Euclidean space. Consider a
hyperplane $H$ such that $0\notin H.$ Let $c=c(H)$ be the foot of the perpendicular dropped from $0$ onto this
hyperplane. Associate with the hyperplane $H$ a point $c^{\ast}=c^{\ast}(H),$ $c^{\ast}=\sigma(c),$ where
$\sigma(c)=\frac{c}{|c|^2}$ is the inversion of the point $c$ with respect to the sphere $S.$

Such a construction can be performed for any hyperplane that does not contain $0.$ The corresponding point
$c^{\ast}(H)$ is uniquely defined by the hyperplane $H.$ Conversely, associate with an arbitrary point $c^{\ast},$
$c^{\ast}\neq 0$ a hyperplane that is perpendicular to the interval $[0,c^{\ast}]$ and that passes through the
point $\sigma(c^{\ast}).$ These constructions set a bijection called polarity between the set of hyperplanes and
the set of points of $n$-dimensional Euclidean space \cite{prep}.

Let us use the polarity for constructing a correspondence
between the set of points of the Euclidean space
and the set of half-spaces.

\begin{center}
{\sc Polar correspondence}
\end{center}

Consider an arbitrary open half-space $P$ with the boundary hyperplane $H$ such that $0\notin H.$ The polarity
associates a point $c^{\ast}$ with the hyperplane $H.$ If $0\in P,$ we color $c^{\ast}$ red; if $0\notin
{\mathrm{cl}}\, P,$ we color $c^{\ast}$ black. Thus, open half-spaces that contain $0$ correspond to red points,
and open half-spaces that do not contain $0$ in their closure correspond to black points.

Conversely, each red (respectively, black) point $c^{\ast},$ $c^{\ast}\neq 0$ corresponds to an open half-space
$P,$ $0\in P$ (respectively, $0\notin {\mathrm{cl}}\, P)$ with the boundary hyperplane $H$ that corresponds to the
point $c^{\ast}$ under the polarity. Thus, the points located at the same place in the space can have two
different colors. Any nonzero point of the space may correspond to one of the two open half-spaces that have a
common boundary hyperplane. This correspondence between the set of open half-spaces and the set of colored points
will be referred to as the polar correspondence.

\begin{center}
{\sc Constructing a system of points \\ from a system of linear inequalities}
\end{center}

Consider the system of linear inequalities
\begin{equation}\label{ks11}
(c_j,x)>b_j, j=1,\ldots ,m
\end{equation}
with nonzero constant terms, and the system ${\cal{D}}$ of $m$ half-spaces, each of which is the set of solutions
to the corresponding inequality. For each half-space of the system ${\cal{D}},$ we construct a colored point that
corresponds to it in polar correspondence. Since, in the general case, the system ${\cal{D}}$ contains the
half-spaces containing $0$ inside, as well as the half-spaces that do not contain $0$ in their closures, the
obtained set ${\cal{D}}'$ of $m$ points contains both red points and black points. Relate to system $(\ref{ks11})$
a system ${\cal{D}}^{\ast}={\cal{D}}' \cup \{0\},$ considering the point $0$ as red. Let ${\cal{A}}$ consist of
all the black points of the system ${\cal{D}}^{\ast}$ and ${\cal{B}}$ consist of all its red points including $0.$
Assume that a ``hyperplane'' of the form $(0,x)=1$ ``separates'' ${\cal{A}}$ from ${\cal{B}}$ if and only if the
subsystem ${\cal{A}}$ contains no points.

{\bf Proposition 1.1 (Polarity of solutions).} {\it A point $x_0$ is a solution to sys\-tem $(\ref{ks11})$ if and
only if the hyperplane $(x_0,x)=1$ corresponding to $x_0$ under po\-la\-ri\-ty strict\-ly se\-pa\-rates the
subsystems of points ${\cal{A}}$ and ${\cal{B}}.$}

{\bf Proof.} The system of colored points ${\cal{D}}^{\ast}={\cal{A}}\cup {\cal{B}}$ corresponding to system
$(\ref{ks11})$ is given by
$${\cal{A}}=\left\{\frac{c_j}{b_j}:j=\overline{1,m}, b_j>0\right\},$$
$${\cal{B}}=\left\{\frac{c_j}{b_j}:j=\overline{1,m}, b_j<0\right\}\cup\{0\}.$$
Indeed, we take an open half-space $P_j=\{x:(c_j,x)>b_j\}$ that is the set of solutions to the $j$th inequality of
system $(\ref{ks11})$ and construct the corresponding colored point. Let $\sigma$ be the inversion with respect to
the unit sphere $S$ centered at $0$ and $\bar{c}_j=\frac{b_j}{|c_j|^2} c_j$ be the foot of the perpendicular
dropped from $0$ onto the boundary $H_j$ of the half-space $P_j.$ The point
$\sigma(\bar{c}_j)=\frac{\bar{c}_j}{|\bar{c}_j|^2}$ coincides with the point $\frac{c_j}{b_j};$ the latter is
colored red if $0\in P_j$ or black if $0\notin {\mathrm{cl}}\, P_j.$ Slightly transform system $(\ref{ks11}).$
Divide the $j$th inequality by the constant term $b_j$ for each $j=1,\ldots,m.$ We obtain the system
$$\left(\frac{c_j}{b_j},x\right )>1,\,\, j\in J_{+}=\{k:b_k>0\},$$
$$\left(\frac{c_j}{b_j},x\right )<1,\,\, j\in J_{-}=\{l:b_l<0\}.$$
Consequently, the point $x_0$ is a solution to system $(\ref{ks11})$ if and only if the hyperplane $(x_0,x)=1$
strictly separates the subsystems of points ${\cal{A}}$ and ${\cal{B}}.$ The proposition is proved.

Consider a
more general case. As the origin, we take an arbitrary vector $z$ that equates none of the inequalities of system
$(\ref{ks11}).$ Analytically, the system of colored points ${\cal{D}}^{\ast}={\cal{A}}\cup {\cal{B}}$
corresponding to system $(\ref{ks11})$ with respect to the origin $z$ is given by
$${\cal{A}}={\cal{A}}(z)=\left\{z+\frac{c_j}{b_j-(c_j,z)}: j=1,\ldots,m, (c_j,z)<b_j\right\},$$
$${\cal{B}}={\cal{B}}(z)=\left\{z+\frac{c_j}{b_j-(c_j,z)}: j=1,\ldots,m, (c_j,z)>b_j\right\}\cup\{z\}.$$
At that same time, the point $x_0$ is a solution to system $(\ref{ks11})$ if and only if the hyperplane
$(x_0-z,x-z)=1$ strictly separates the subsystems of points ${\cal{A}}$ and ${\cal{B}}.$ To prove this, it is
sufficient to make the change of variables $y=x-z$ and apply Proposition 1.1 to the system
$$(c_j,y)>b_j-(c_j,z), j=1,\ldots,m,$$
as well as to the system of points ${\cal{D}}^{\ast}_0={\cal{A}}_0\cup {\cal{B}}_0$ corresponding to this system
with respect to $0,$ where ${\cal{A}}_0={\cal{A}}-z$ and ${\cal{B}}_0={\cal{B}}-z.$ Since the solution $y_0$ to
this system corresponds to the solution $y_0+z$ of system $(\ref{ks11}),$ without loss of generality, we may
assume that system $(\ref{ks11})$ corresponds to the system of points ${\cal{D}}^{\ast}_0$ with respect to the
origin $z.$ Formulate also an analog of Proposition 1.1 for an arbitrary system of nonstrict linear inequalities
\begin{equation}\label{ks12}
(c_j,x)\geq b_j, j=1,\ldots,m
\end{equation}
with nonzero constant terms. Consider the system of points ${\cal{D}}^{\ast}={\cal{A}}\cup{\cal{B}}$ that is
constructed on the basis of the system of strict linear inequalities obtained from system $(\ref{ks12})$ by
replacing the $\geq$ sign with the $>$ sign in each inequality.

{\bf Proposition 1.2.} {\it A point $x_0$ is a solution to system $(\ref{ks12})$ if and only if the hyperplane
$(x_0,x)=1$ corresponding to this point in polarity separates the sub\-sys\-tems of points ${\cal{A}}$ and
${\cal{B}}$ and does not pass through $0.$}

Consequently, finding a solution to a system of linear inequalities can be represented as constructing a
hyperplane separating two point sets corresponding to this system.

\begin{center}
{\sc Inconsistent systems of linear inequalities}
\end{center}

Consider an inconsistent system of linear inequalities
\begin{equation}\label{ks13}
(c_j,x)>b_j, j=\overline{1,m}
\end{equation}
with nonzero constant terms. Find the system of colored points ${\cal{D}}^{\ast}={\cal{A}}\cup{\cal{B}}$ that
corresponds to this system with respect to $0.$

We say that an open half-space $P$, $0\notin{\mathrm{cl}}\,P$ votes for a point $a\in{\cal{A}}$ if $a\in P$ and
votes for a point $b\in{\cal{B}}$ if $b\notin {\mathrm{cl}}\,P.$ Moreover, $P$ votes for a subsystem of points if
$P$ votes for each point of this subsystem; $P$ votes against a point if $P$ does not vote for this point.

{\bf Definition 1.1.} A committee of the system of colored points ${\cal{D}}^{\ast}={\cal{A}}\cup{\cal{B}}$ is a
finite collection (with possible repetitions) of open half-spaces not containing $0$ in their closures such that,
if $d$ is any point of system ${\cal{D}}^{\ast},$ then a majority (dependent of $d)$ of half-spaces of this
collection votes for $d.$

{\bf Proposition 1.3.} {\it The set $K=\{x^1,\ldots,x^q\}$ is a committee of system $(\ref{ks13})$ if and only if
the set $K^{\ast}=\{P_1,\ldots,P_q\}$ consisting of the half-spaces $P_i=\{x:(x^i,x)>1\},$ where $i=1,\ldots,q,$
is a committee of the system of points ${\cal{D}}^{\ast}={\cal{A}}\cup{\cal{B}}.$}

{\bf Proof.} Consider an arbitrary committee $K=\{x^1,\ldots,x^q\}$ of system $(\ref{ks13})$ and the point
$c^{\ast}_j=\frac{c_j}{b_j}\in {\cal{D}}^{\ast}$ corresponding to the $j$th inequality of system $(\ref{ks13}).$
If a member $x^{i_0},$ $1\leq i_0\leq q,$ of $K$ satisfies this inequality, then $(\frac{c_j}{b_j},x^{i_0})>1$
$b_j>0$ and $(\frac{c_j}{b_j},x^{i_0})<1$ for $b_j<0.$ Thus, the half-space $P_{i_0}$ votes for the point
$c^{\ast}_j.$ Since the majority of committee $K$ members satisfies the $j$th inequality, the majority of
half-spaces in the set $K^{\ast}=\{P_1,\ldots,P_q\}$ votes for the point $c^{\ast}_j.$ Consequently, $K^{\ast}$ is
a committee of the system of points ${\cal{D}}^{\ast}.$ The same arguments in the reverse order finish the proof.

For brevity, we call an ``MCS of system $(\ref{ks13})$'' the maximal by inclusion con\-sis\-tent subsystem of
system $(\ref{ks13}).$

{\bf Proposition 1.4.} {\it Let ${\cal{D}}^{\ast}={\cal{A}}\cup{\cal{B}}$ be a system of points that
cor\-res\-ponds to system $(\ref{ks13})$ with respect to the origin $z\in \RR^n.$ Then,

(a) if $z$ satisfies a certain $MCS$ of system $(\ref{ks13})$ then ${\cal{A}}\subset {\mathrm{conv}}\,{\cal{B}};$

(b) if $z$ is a solution to some MCS of the system
$$(c_j,x)<b_j,\, j=1,\ldots,m$$
then ${\cal{B}}\subset {\mathrm{conv}}\,({\cal{A}}\cup\{0\}).$}

In case (a), Proposition 1.4 is easy to prove using the theorem on strict se\-pa\-ra\-bi\-li\-ty of a point from a
closed convex set. In case (b), the proposition follows from (a).\clearpage

\section{PROCEDURE FOR FINDING\\ THE MEMBERS
OF A COMMITTEE}

\begin{center}
{\sc{Marked MCSs of inconsistent system}}
\end{center}

Consider an inconsistent system of linear inequalities
\begin{equation}\label{ks21}
(c_j,h)>b_j,\, j=1,\ldots,m,\, c_j,h\in\RR^2,
\end{equation}
that has a committee. Let $|c_j|=1,$ $j=1,\ldots,m.$ Assume that system $(\ref{ks21})$ does not contain identical
inequalities.

{\bf Definition 2.1.} An MCS $T$ of system $(\ref{ks21})$ is called
{\it marked} if it contains a pair of inequalities, called {\it determining},
such that any MCS of system $(\ref{ks21})$ containing
this pair coincides with $T,$ and any solution to the determining
pair of the inequalities sufficiently close to an
equating element for the inequalities of this pair satisfies
the subsystem $T$.

It can be shown that any marked MCS of system $(\ref{ks21})$ includes only one de\-ter\-mi\-ning pair of
inequalities.

{\bf Proposition 2.1.} {\it Let $T$ be an arbitrarily marked
MCS of system $(\ref{ks21})$ and $S$ be the determining pair of
the inequalities that belongs to $T.$ Then, $T$ consists of all
the inequalities of system $(\ref{ks21})$ forming, together with
two inequalities from $S,$ a consistent subsystem.}

The proposition follows from Definition 2.1.

{\bf Proposition 2.2.} {\it Let $K$ be a committee of system
$(\ref{ks21})$ such that each element of $K$ is a solution to some
MCS of system $(\ref{ks21}).$ Then, any marked MCS’s of this
system has a solution entering $K$ with a positive multiplicity.}

This proposition follows from the fact that any pair of inequalities of system $(\ref{ks21})$ has a solution
belonging to a committee of this system. In particular, Pro\-po\-si\-ti\-on 2.2 implies that the number of marked
MCS’s of system $(\ref{ks21})$ is not greater than the number of elements of its minimal committee.

Let the half-plane $D_j$ be the set of solutions to the $j$th
inequality of system $(\ref{ks21}),$ and ${\cal{D}}=\{D_j\}_{j=1}^m$
be a system
of half-planes. Below, a ray with its vertex removed will
also be referred to as a ray. We call a nonempty intersection
of any two closed (respectively, two open) half-planes
with nonparallel borders a cone whose vertex is the intersection
point of the borders of these two half-planes.

{\bf Procedure for finding a marked MCS.} Consider a half-plane $D^{(0)}\in {\cal{D}}$ and a point $h^{(0)}$ on
its border $l^{(0)}$ that coincides with one of the two extreme points of the intersection $l^{(0)}$ with the
border lines of the other half-planes of the system ${\cal{D}}.$ On the line $l^{(0)},$ take a ray $r_0$
containing both extreme points with the origin at the point $h^{(0)}.$ If these points coincide, we take any of
the two rays lying on the line $l^{(0)}$ with the origin at the point $h^{(0)}.$ Consider the following process.

The intersection $\bigcap\{D\cap l^{(0)}:D\in{\cal{D}},D\cap r_0 \mbox{is a ray}\}$ coincides with a ray $D'\cap
l^{(0)}$,$D'\in {\cal{D}}$, and the intersection $\bigcap\{D\cap D^{(0)}:D\in{\cal{D}},D\cap l^{(0)}=D'\cap
l^{(0)}\}$ coincides with a cone $D^{(1)}\cap D^{(0)}$ whose vertex is denoted by $h^{(1)}$; $h^{(1)}=l^{(0)}\cap
l^{(1)}$, where $l^{(1)}$ is the border of the half-plane $D^{(1)}.$ Let $r_1=D^{(0)}\cap l^{(1)}.$

Similarly, the intersection $\bigcap\{D\cap l^{(1)}:D\in{\cal{D}},D\cap r_1 \mbox{is a ray}\}$ coincides with a
ray $D''\cap l^{(1)},$ $D''\in {\cal{D}},$ and the intersection $\bigcap\{D\cap D^{(1)}:D\in{\cal{D}},D\cap
l^{(1)}=D''\cap l^{(1)}\}$, coincides with a cone $D^{(2)}\cap D^{(1)}$, $D^{(2)}\in {\cal{D}}$. Let $h^{(2)}$ be
the vertex of this cone, $h^{(2)}=l^{(1)}\cap l^{(2)},$ where $l^{(2)}$ is the border of the half-plane $D^{(2)}.$
If $h^{(1)}=h^{(2)},$ then the procedure is terminated. In the case of $h^{(1)}\neq h^{(2)},$ we make the same
construction for the line $l^{(2)}$ and the ray $r_2=D^{(1)}\cap l^{(2)}$. Then, we find a half-plane $D^{(3)}$
and the vertex $h^{(3)}$ of the cone $D^{(3)}\cap D^{(2)}$, $h^{(3)}=l^{(3)}\cap l^{(2)}$, where $l^{(3)}$ is the
border of $D^{(3)}$, and set $r_3=D^{(2)}\cap l^{(3)}$. We go on with this construction process until the equality
$h^{(k_0-1)}=h^{(k_0)}$ is satisfied for some $k_0$, $k_0\geq 2$. Let ${\cal{D}}'=\{D^{(s)}\}_{s=0}^{k_0}.$

We choose a point $\bar{h}\in D^{(k_0-1)}\cap D^{(k_0)}$ that belongs
to so small a neighborhood of the point  $h^{(k_0)}$ that
$\bar{h}\in\bigcap\{D\in{\cal{D}}:h^{(k_0)}\in D\}$.
Each of the half-planes $D^{(k_0-1)}$ and $D^{(k_0)}$
is the set of solutions to an inequality
of system $(\ref{ks21})$. Denote by $S$ the subsystem of system
$(\ref{ks21})$ composed of these two inequalities. Consider an
arbitrary inequality of system $(\ref{ks21})$ that forms a consistent
subsystem with the two inequalities from $S.$ Compose
the subsystem consisting of all such inequalities.
Denote it by $T.$ The procedure is complete.

{\bf Theorem 2.1.} {\it The introduced procedure is correct. The subsystem $T$ co\-in\-cides with a marked MCS of
system $(\ref{ks21}).$ Two inequalities of $S$ form a determining pair in $T,$ and the vector $\bar{h}$ is a
solution to the subsystem $T.$}

{\bf Proof.} Let us prove the correctness of the procedure. Since system $(\ref{ks21})$ is in\-con\-sis\-tent and
any its subsystem of two inequalities is consistent, the rank of system $(\ref{ks21})$ equals two. Consequently,
there exists a half- plane $D\in{\cal{D}}$ such that its border is not parallel to the line $l^{(0)}$. Since we
can choose two extreme points (possible coinciding) among all the points of intersection of the line $l^{(0)}$
with the borders of other half-planes of ${\cal{D}}$, the choice of the point $h^{(0)}$ and the ray $r_0$ is
possible.

Let ${\cal{D}}_1=\{D\in{\cal{D}}:D\cap l^{(0)}\mbox{is a ray}\}$. We show that
there exists a half-plane $D'\in{\cal{D}}_1$ such that the ray
$D'\cap l^{(0)}$ is codirectional with the ray $r_0$. By contradiction,
let, for any half-plane $D\in{\cal{D}}_1$, the rays $D\cap
l_0$
and $r_0$ be counterdirectional. Consider the system
${\cal{D}}_2=\{D\in {\cal{D}}: {\mathrm{bd}}\,D\parallel
l^{(0)}\}$, where the $\parallel$ sign denotes the parallelism
of two lines. Since the intersection of any two
half-planes of the system ${\cal{D}}_2$ is nonempty, by the Helly
theorem there exists a point $h'\in
C=\bigcap\{D: D\in{\cal{D}}_2\}$.
Consider the ray $\bar{r}$ with the origin at the point $h'$ that is
counterdirectional to $r_0$. Clearly, $\bar{r}\subset C$. Since the intersection
$\bigcap\{D\cap \bar{r}:D\in{\cal{D}}_1\}$ is a ray, some point $\bar{h}\in \bar{r}$
belongs to the intersection $\bigcap\{D: D\in{\cal{D}}\}$, which is
impossible due to the inconsistency of system $(\ref{ks21})$.

Therefore, there exists a half-plane $\bar{D}\in{\cal{D}}_1$ such
that the rays $\bar{D}\cap l^{(0)}$ and $r_0$ are codirectional and, consequently,
$\bar{D}\cap l^{(0)}\in\{D\cap
l^{(0)}:D\in{\cal{D}},D\cap r_0\mbox{is a ray}\}$.
Since the intersection
$\bigcap\{D\cap l^{(0)}:D\in{\cal{D}},D\cap r_0\mbox{is a ray}\}$
coincides with a ray $D'\cap l^{(0)}$, $D'\in{\cal{D}},$ we have
$D'\cap D^{(0)}\in\{D\cap
D^{(0)}:D\in{\cal{D}},D\cap l^{(0)}=D'\cap l^{(0)}\}$.

Since the intersection $\bigcap\{D\cap D^{(0)}:D\in{\cal{D}},D\cap l^{(0)}=D'\cap l^{(0)}\}$ coincides with a cone
$F\cap D^{(0)}$, $F\in {\cal{D}}$, the choice of the half-plane $D^{(1)}$ is possible. It can be proved in a
similar way that the choice of the half-plane $D^{(k)}$ is also possible for any $k>1$. Show that $D^{(k-1)}\cap
D^{(k)}\subseteq D^{(k-2)}\cap D^{(k-1)}$ for any $k\geq 2$. We prove this inclusion for $k=2$ since, for $k>2$,
the proof is similar. If $h^{(2)}=h^{(1)}$, then $r_1=D^{(2)}\cap l^{(1)}$ in view of the equality
\begin{equation}\label{ks27}
D^{(2)}\cap l^{(1)}=\bigcap\{D\cap l^{(1)}:D\in{\cal{D}},D\cap r_1 - \mbox{луч}\}.
\end{equation}
Therefore, $D^{(0)}\cap D^{(1)}\in\{D\cap D^{(1)}:D\in{\cal{D}},D\cap l^{(1)}=D^{(2)}\cap l^{(1)}\}$. Then,
$D^{(1)}\cap D^{(2)}\subseteq D^{(0)}\cap D^{(1)}$ owing to the equality
\begin{equation}\label{ks26}
D^{(2)}\cap D^{(1)}=\bigcap\{D\cap D^{(1)}:D\in{\cal{D}},D\cap l^{(1)}=D^{(2)}\cap l^{(1)}\}.
\end{equation}
Consider the case of $h^{(2)}\neq h^{(1)}$. In this case, $D^{(2)}\cap l^{(1)}\subset r_1$; consequently,
$h^{(1)}\notin {\mathrm{cl}}\, D^{(2)}$ and $h^{(2)}\in D^{(0)}$. If the half- plane $E\in{\cal{D}}$ is such that
the rays $E\cap l^{(0)}$ and $D^{(1)}\cap l^{(0)}$ are codirectional, then $E\cap l^{(0)}\in\{D\cap
l^{(0)}:D\in{\cal{D}},D\cap D^{(1)}\cap l^{(0)}\mbox{is a ray}\}$, Moreover, since the ray $D^{(1)}\cap l^{(0)}$
is codirectional to the ray $r_0$, for any $D\in{\cal{D}}$, the intersection $D\cap r_0$ is a ray if and only if
the intersection $D\cap D^{(1)}\cap l^{(0)}$ is a ray. This implies
\begin{equation}\label{ks28}
D^{(1)}\cap l^{(0)}=\bigcap\{D\cap l^{(0)}:D\in{\cal{D}},D\cap D^{(1)}\cap l^{(0)} - \mbox{луч}\},
\end{equation}
and, consequently, $D^{(1)}\cap l^{(0)}\subseteq E\cap l^{(0)}$. Therefore, ${\mathrm{cl}}\,D^{(2)}\cap
{\mathrm{cl}}\, D^{(1)}\cap l^{(0)}=\varnothing$. Then, ${\mathrm{cl}}\,D^{(1)}\cap l^{(2)}\subset D^{(0)}$.
Hence, each of the two border rays of the cone ${\mathrm{cl}}(D^{(1)}\cap D^{(2)})$ belongs to the half-plane
$D^{(0)}$. Then, ${\mathrm{cl}}(D^{(1)}\cap D^{(2)})\subset D^{(0)}$, which implies $D^{(1)}\cap D^{(2)}\subset
D^{(0)}\cap D^{(1)}$.

It can be similarly proved that ${\mathrm{cl}}(D^{(k-1)}\cap D^{(k)})\subset D^{(k-2)}$ for $h^{(k)}\neq
h^{(k-1)}$, $k\geq 2$. Then, for $h^{(k)}\neq h^{(k-1)}$, $k\geq 2$, the sequence $\{D^{(s)}\}_{s=0}^k\subseteq
{\cal{D}}$ consists of $k+1$ different half-planes. Therefore, $h^{(k)}=h^{(k-1)}$ for a positive integer $k=k_0$,
$k_0\geq 2$. Since, evidently, $D^{(k_0-1)}\cap D^{(k_0)}\neq\varnothing$, the choice of the point $\bar{h}$ and
the subsystem $T$ is possible and, consequently, the introduced procedure is correct.

We show that $\bar{h}\in \bigcap\{D\in{\cal{D}}:D\cap D^{(k_0-1)}\cap D^{(k_0)}\neq\varnothing\}$. Let $k_0=2$ for
simplicity. For $k_0>2$, the proof is analogous. Consider a half-plane $E\in{\cal{D}}$ with the border $l$ such
that $E\cap D^{(1)}\cap D^{(2)}\neq\varnothing$ and $h^{(2)}\in l$. Let us show that $D^{(1)}\cap D^{(2)}\subset
E$. Since $h^{(1)}=h^{(2)}$, we have $h^{(2)}=l^{(0)}\cap l^{(1)}\cap l^{(2)}$ and, consequently, $D^{(2)}\cap
l^{(1)}=r_1$ by virtue of equality $(\ref{ks27})$. Since $D^{(1)}\cap D^{(2)}\subseteq D^{(0)}\cap D^{(1)}$, we
have $E\cap D^{(0)}\cap D^{(1)}\neq\varnothing$ The inclusion $h^{(2)}\in l$ implies that $E$ contains at least
one of the border rays of the cone $D^{(0)}\cap D^{(1)}$. Therefore, one of two cases occurs: either $D^{(2)}\cap
l^{(1)}=E\cap l^{(1)}$ or $D^{(1)}\cap l^{(0)}=E\cap l^{(0)}$. In the first case, $D^{(1)}\cap D^{(2)}\subseteq
D^{(1)}\cap E$ in view of equality $(\ref{ks26})$. This implies $D^{(1)}\cap D^{(2)}\subset E$. In the second
case, $D^{(0)}\cap D^{(1)}\subseteq D^{(0)}\cap E$, because $D^{(1)}\cap D^{(0)}=\bigcap\{D\cap
D^{(0)}:D\in{\cal{D}},D\cap l^{(0)}=D^{(1)}\cap l^{(0)}\}$. However, in this case, $D^{(0)}\cap D^{(1)}\subset E$,
whence $D^{(1)}\cap D^{(2)}\subset E$.

Consider an arbitrary half-plane $E\in{\cal{D}}$ with the border $l$ such that $E\cap D^{(1)}\cap
D^{(2)}\neq\varnothing$ and $h^{(2)}\notin l$. We show that $h^{(2)}\in E$. If none of the two intersections
$E\cap D^{(2)}\cap l^{(1)}$ and $E\cap D^{(1)}\cap l^{(2)}$ is a ray, then the set ${\mathrm{cl}}\,(E\cap
D^{(1)}\cap D^{(2)})$ is a triangle with a vertex $h^{(2)}$. Thus, $h^{(2)}\in E$. Let one of the intersections
$E\cap D^{(2)}\cap l^{(1)}$ and $E\cap D^{(1)}\cap l^{(2)}$ be a ray. Since $D^{(1)}\cap D^{(2)}\subset D^{(0)}$,
the set $E\cap D^{(0)}\cap D^{(1)}$ is unbounded. Since $h^{(1)}=h^{(2)}$, we have $D^{(2)}\cap l^{(1)}=r_1$.
Consequently, at least one of the intersections $E\cap D^{(2)}\cap l^{(1)}$ and $E\cap D^{(1)}\cap l^{(0)}$ is a
ray. If $E\cap D^{(2)}\cap l^{(1)}$ is a ray, then $h^{(2)}\in E$ in view of equality $(\ref{ks27})$ and taking
into account $h^{(2)}\notin l$. If $E\cap D^{(1)}\cap l^{(0)}$ is a ray, then $h^{(2)}\in E$ by virtue of equality
$(\ref{ks28})$. Thus, $h^{(2)}\in\bigcap\{D\in{\cal{D}}:D\cap D^{(1)}\cap D^{(2)}\neq\varnothing, h^{(2)}\notin
{\mathrm{bd}}\,D\}$. From the proved above, it follows that $D^{(1)}\cap D^{(2)}\subseteq\bigcap\{D\in{\cal{D}}:
D\cap D^{(1)}\cap D^{(2)}\neq\varnothing, h^{(2)}\in{\mathrm{bd}}\,D\}$. Then, $\bar{h}\in
M=\bigcap\{D\in{\cal{D}}:D\cap D^{(1)}\cap D^{(2)}\neq\varnothing\}$ by construction of the point $\bar{h}$.
Moreover, any element of the intersection $D^{(1)}\cap D^{(2)}\cap \bigcap\{D\in{\cal{D}}:D\cap D^{(1)}\cap
D^{(2)}\neq\varnothing, h^{(2)}\notin {\mathrm{bd}}\,D\}$ belongs to $M$.

Since the set of solutions to any inequality of the subsystem $T$ is a half-plane $D\in{\cal{D}}$ such that $D\cap
D^{(k_0-1)}\cap D^{(k_0)}\neq\varnothing$ it follows that $\bar{h}$ is a solution to the subsystem $T$. Hence, $T$
is an MCS of system $(\ref{ks21})$. The half-planes $D^{(k_0-1)}$ and $D^{(k_0)}$ are the sets of solutions to the
inequalities, which form the subsystem $S.$ Consequently, the subsystem $T$ is a marked MCS of system
$(\ref{ks21})$, and $S$ is the determining pair of inequalities in $T.$ The theorem is proved.

Consider an arbitrary half-plane $D^{(0)}\in{\cal{D}}$ and a ray
$r_0$ on its border $l^{(0)}$ containing both extreme points of
the intersection of $l^{(0)}$ with the borders of the other half-
planes of the system ${\cal{D}}$ and having one of these extreme
points as the vertex. Denote this vertex by $h^{(0)}$. Applying
the procedure to the given half-plane $D^{(0)}$, the line
$l^{(0)}$, and the ray $r_0$, we obtain a solution $\bar{h}$ to some
marked MCS $T$ of system  $(\ref{ks21})$. Let ${\cal{D}}'=\{D^{(s)}\}_{s=0}^{k_0}\subset {\cal{D}}$
be the sequence of half-planes formed by the procedure.
Consider an arbitrary inequality of system $(\ref{ks21})$
such that the set of its solutions (that is a half-plane)
belongs to ${\cal{D}}'$. Compose a subsystem of all these inequalities
and denote it by $T'$.

{\bf Corollary 2.1.1.} {\it The subsystem $T$ contains the subsystem
$T'.$}

Further, for brevity, we consider the application of
the procedure to the half-plane $D^{(0)}\in{\cal{D}}$ and the ray $r_0$
on its border $l^{(0)}$ under the assumption that the ray $r_0$
contains both extreme points of the intersection $l^{(0)}$ with
the borders of other half-planes of the system ${\cal{D}}$, and
that it has its origin at one of these extreme points.

{\bf Proposition 2.3.} {\it Let $T_0$ be an arbitrary marked
MCS of system $(\ref{ks21})$ with the determining pair of inequalities
$S_0.$ Then, there exist a half-plane $D^{(0)}\in {\cal{D}}$ and
a ray $r_0$ on its border such that $T_0=T$ and $S_0=S,$ where
$T$ and $S$ are the subsystems obtained by applying the
procedure to the half-plane $D^{(0)}$ and the ray $r_0.$ The
obtained vector $\bar{h}$ is a solution to the subsystem $T_0.$}

{\bf Proof.} Let the half-planes $E$ and $E'$ be the sets of
solutions of two inequalities composing the subsystem
$S_0.$ Set $D^{(0)}:=E.$ Let $T$ and $S$ be the subsystems
obtained by applying the procedure to the half-plane
$D^{(0)}$ and the ray $r_0$ on
its border $l^{(0)}$ codirectional to the ray
$E'\cap l^{(0)}$. Following this procedure and applying
Proposition 2.1 and Definition 2.1, we can show that
this procedure forms a sequence ${\cal{D}}'=\{D^{(s)}\}_{s=0}^2\subset {\cal{D}}$
such that $D^{(1)}=E'$ and $D^{(2)}=E$. Therefore, $S_0=S$ and,
consequently, $T_0=T.$ The proposition is proved.

Denote by $l_j$ the boundary line of the half-plane $D_j,\,j=1,\ldots,m.$

{\bf Corollary 2.3.1.} {\it The $i$th and $j$th inequalities of system
$(\ref{ks21})$, $i\neq j$, form a determining pair for some
marked MCS if and only if $(c_i,\bar{h})=b_i, (c_j,\bar{h})=b_j$ for
some $\bar{h}\in{\mathbf{R}}^2$ and the following equalities hold:}
\begin{equation}\label{ks221}
D_i\cap l_j=\bigcap\{D\cap l_j:D\in{\cal{D}},D\cap D_i\cap l_j\; \mbox{is a ray}\},
\end{equation}
\begin{equation}\label{ks222}
D_j\cap l_i=\bigcap\{D\cap l_i:D\in{\cal{D}},D\cap D_j\cap l_i\; \mbox{is a ray}\},
\end{equation}
$$D_i\cap D_j=\bigcap\{D\cap D_j:D\in{\cal{D}},D\cap l_j=D_i\cap
l_j\}=$$
\begin{equation}\label{ks223}
=\bigcap\{D\cap D_i:D\in{\cal{D}},D\cap l_i=D_j\cap l_i\}.
\end{equation}

\begin{center}
{\sc{Exact upper and lower estimates\\ for the number of marked MCSs of inconsistent system}}
\end{center}

Along with system $(\ref{ks21})$, consider the corresponding
system
\begin{equation}\label{ks23}
(c_j,h)>0, j=\overline{1,m}.
\end{equation}

{\bf Proposition 2.4.} {\it Let $n_0$ be the number of marked
MCS’s of system $(\ref{ks21})$. Then, $n_0\geq 3$. If there exists a
committee of system $(\ref{ks23})$, then $n_0\leq q_{H}$, where $q_H$ is the
number of elements of the minimal committee of system
$(\ref{ks23})$.}

{\bf Proof.} Let us prove that $n_0\geq 3$. Let $D^{(0)}\in
{\cal{D}}$ be an
arbitrary half-plane. Applying the procedure to the half-
plane $D^{(0)}$ and an arbitrary ray on its border, we find a
marked MCS $T_1$ of system $(\ref{ks21})$ and the determining
pair of inequalities $S_1$ belonging to $T_1.$ Let the half-
planes $E$ and $E'$ be the sets of solutions to the inequalities
from $S_1.$ Consider the marked MCS $T_2$ obtained by
applying the procedure to the half-plane $D^{(0)}=E$ and to
the ray $r_0$ on its border $l^{(0)}={\mathrm{bd}}\,E$, which is counterdirectional
to the ray $E'\cap l^{(0)}$. We show that $T_1\neq T_2.$ Following
the procedure for finding the subsystem $T_2,$ we
consider the intersection $\bigcap\{D\cap
l^{(0)}:D\in{\cal{D}},D\cap r_0\mbox{is a ray}\}$ that coincides
with a ray $D'\cap l^{(0)}$, $D'\in {\cal{D}}$.
Since $T_1$ does not coincide with system $(\ref{ks21})$, by Proposition
2.1, there exists a half-plane $\bar{D}\in {\cal{D}}$ such that
$\bar{D}\cap E\cap E'=\varnothing.$ Since at the same time $\bar{D}\cap E\neq\varnothing$
and $\bar{D}\cap
E'\neq\varnothing$, the intersection $\bar{r}=\bar{D}\cap l^{(0)}$ is a ray.
Moreover, the rays $\bar{r}$ and $E'\cap l^{(0)}$ do not intersect; consequently,
the rays $r_0$ and $\bar{r}$ are codirectional. Thus, the
intersection $\bar{D}\cap r_0$ is a ray. Since the half-planes $E$ and
$E'$ are the sets of solutions to the inequalities of the
determining pair in the marked MCS, any element of
the cone $E\cap E'$ lying in a sufficiently small neighborhood
of its vertex belongs to the intersection$\bigcap\{D\in{\cal{D}}:D\cap
E\cap E'\neq\varnothing\}$. Since $\bar{r}\in\{D\cap
l^{(0)}:D\in{\cal{D}},D\cap r_0\mbox{is a ray}\}$, it follows that $D'\cap
l^{(0)}\subseteq \bar{r}$, whence $D'\cap E\cap E'=\varnothing$.
Following the procedure, consider the intersection
$\bigcap\{D\cap
D^{(0)}:D\in{\cal{D}},D\cap l^{(0)}=D'\cap l^{(0)}\}$ coinciding
with some cone $D^{(1)}\cap D^{(0)}$, $D^{(1)}\in {\cal{D}}$. By Corollary
2.1.1, the inequality of system $(\ref{ks21})$ whose set of solutions
is $D^{(1)}$ belongs to the subsystem $T_2.$ Taking into
account $D^{(0)}=E$, we have $D^{(1)}\cap E\subseteq D'\cap E$, whence
$D^{(1)}\cap
E\cap E'=\varnothing$. Thus, this inequality does not belong
to the subsystem $T_1$, i.e., $T1\neq T2.$

We set $D^{(0)}:=E'$ and consider the marked MCS $T_3$ obtained by applying the procedure to the half-plane
$D^{(0)}$ and the ray $r_0$ on its border $l^{(0)}$ that is counterdi\-rec\-tio\-nal to the ray $E\cap l^{(0)}$.
Similarly, it can be shown that $T_1\neq T_3.$ Let us prove that $T_2\neq T_3.$ By Corollary 2.1.1, the inequality
of system $(\ref{ks21})$ whose set of solutions is the half-plane $E$ belongs to the subsystem $T_2,$ and the
inequality of this system whose set of solutions is $E'$ belongs to $T_3.$ In the case of $T_2=T_3$, we obtain
that the pair of inequalities $S_1$ belongs to each of the subsystems $T_2$ and $T_3$, and, consequently,
$T_1=T_2=T_3$, what is impossible by the proved above. Thus, the subsystems $T_1$, $T_2$, and $T_3$ are different;
hence, $n_0\geq 3$.

Suppose that there exists a committee of system $(\ref{ks23})$. We show that $n_0\leq q_H$. As it was proved in
\cite{mazurov}, the number $q$ of members of a minimal committee of system $(\ref{ks21})$ is not greater than
$q_H$. However, $n_0\leq q$, that is $n_0\leq q_H$. The proposition is proved.

The upper and lower estimates for the number $n_0$ given in Proposition 2.4 are sharp.

{\bf Example 1.} Let $q$ be an odd integer, $q>1$. Consider
a regular $q$-gon such that the set $V$ of its vertices lies on
the unit circle centered at $0.$ Consider the system of $q$
linear inequalities
\begin{equation}\label{ks25}
(c,h)>1, c\in V, h\in{\mathbf{R}}^2.
\end{equation}
It can be shown that the number of members of a minimal
committee of this system is equal to the same number
for the system
$$(c,h)>0,c\in V,$$
and equals $q.$ At the same time, the number of marked
MCSes of system $(\ref{ks25})$ is also equal to $q.$

Below $[x]$ denotes the integer part of the number $x,$
and $k\mod l$ denotes the remainder from division of $k$ by
$l,$ where $k$ and $l$ are integers.

{\bf Example 2.} Let $q$ be an odd integer, $q>1.$ We partition
the unit circle $C$ centered at $0$ into six arcs of
length $\pi/3$ and choose three pairwise disjoint arcs $\gamma_1$, $\gamma_2$,
and $\gamma_3$ among them. Let $p=[\frac{q+3}{6}]$
and $r=\frac{q+3}{2}\mod 3$.
For each $i=1,2,3$, consider a finite subset of points ${\cal{A}}_i$
on the arc $\gamma_i$ that does not contain the endpoints of this
arc, $|{\cal{A}}_i|=p$,
$i=\overline{1,3}$ in the case where $r=0$ and $|{\cal{A}}_i|=p+1$
$i=\overline{1,r}$, $|{\cal{A}}_i|=p$, $i=\overline{r+1,3},$ in the case
where $r>0$. Let ${\cal{A}}=\bigcup\limits_{i=1}^3
{\cal{A}}_i$. Obviously, $|{\cal{A}}|=3p+r=\frac{q+3}{2}.$

Consider two different points $a$ and $a'$ from ${\cal{A}}_1$ such that the shortest arc $\omega$ of the circle
$C$ connecting them contains no points of the set ${\cal{A}}_1$ other than $a$ and $a'$. Take a point $b$ on the
arc $\omega$ that is equidistant from its endpoints. Compose a finite set ${\cal{B}}_1\subset\gamma_1$ of the
points $b=b(a,a')$ for all the pairs of different points $a$ and $a'$ of the set ${\cal{A}}_1$ satisfying this
condition. Construct in a similar way finite subsets of points ${\cal{B}}_2\subset\gamma_2$ and
${\cal{B}}_3\subset\gamma_3$. It is easy to see that $|{\cal{B}}_i|=|{\cal{A}}_i|-1$, $i=\overline{1,3}$. Let
${\cal{B}}=\bigcup\limits_{i=1}^3 {\cal{B}}_i$. Obviously, $|{\cal{B}}|=\frac{q-3}{2}$. Consider the system of $q$
linear inequalities
$$(a,h)>1,\,a\in{\cal{A}},\,\,
(b,h)<1,\,b\in{\cal{B}}.$$ It can be shown that the number of members of its minimal committee is equal to $q$ and
the number of marked MCS’s is equal to three.\clearpage

\begin{center}
{\sc{Consistent systems}}
\end{center}

Consider a consistent system of linear inequalities
\begin{equation}\label{ks24}
(c_j,h)>b_j,\, j=1,\ldots,m, c_j,h\in \mathbf{R}^2.
\end{equation}
Let $|c_j|=1$, $j=\overline{1,m}$. Suppose that system $(\ref{ks24})$ does not contain identical
ine\-qua\-li\-ties and has rank two. Consider the system of the half-planes${\cal{D}}=\{D_j\}_{j=1}^m$ where $D_j$
is the set of solutions to the $j$th inequality of system $(\ref{ks24})$.

{\bf Proposition 2.5.} {\it For any half-plane $D^{(0)}\in{\cal{D}}$, there
exists a ray $r_0$ on its border such that the point $\bar{h}$
obtained by applying the procedure to the half-plane
$D^{(0)}$ and to the ray $r_0$ is a solution to system $(\ref{ks24})$.}

{\bf Proof.} Let $l^{(0)}$ be the border of the half-plane $D^{(0)}$. Since the rank of system $(\ref{ks24})$
equals two, there exists a half-plane $E\in {\cal{D}}$ whose border is not parallel to the line $l^{(0)}$. Let
$r_0$ be a codirectional ray to the ray $\bar{r}=E\cap l^{(0)}$ on the line $l^{(0)}$ that contains two extreme
points of the intersection of $l^{(0)}$ with the borders of the other half- planes of the system ${\cal{D}}$ and
that has a vertex at one of the extreme points. It is not difficult to show that the choice of the half-plane
$D^{(k)}$ is possible for each $k\geq 1$.

Following the arguments of the proof of Theorem
2.1, we can similarly show that the procedure forms a
finite sequence ${\cal{D}}'=\{D^{(s)}\}_{s=0}^{k_0}\subseteq {\cal{D}}$, and the point $\bar{h}$
obtained by applying the procedure belongs to the
intersection $\bigcap\{D\in{\cal{D}}:D\cap D^{(k_0-1)}\cap
D^{(k_0)}\neq\varnothing\}$.
Since system $(\ref{ks24})$ is consistent, the vector $\bar{h}$ is obviously
its solution. The proposition is proved.

\begin{center}
{\sc{Finding MCS\\ containing a given subsystem}}
\end{center}

Let ${\cal{J}}_0$ be an arbitrary consistent subsystem of system $(\ref{ks21})$ of rank 2. Pro\-po\-si\-tion 2.5
implies the following method for finding an MCS of system $(\ref{ks21})$ containing the subsystem ${\cal{J}}_0$.
It is based on the use of the procedure. Consider a system of half-planes ${\cal{D}}_0\subset{\cal{D}}$ that are
the sets of solutions to the inequalities of the subsystem ${\cal{J}}_0$ and a vector $h_0$ satisfying the
subsystem ${\cal{J}}_0$.

{\bf Algorithm for finding an MCS.} Let ${\cal{D}}'={\cal{D}}_0$ and ${\cal{D}}''={\cal{D}}\backslash
{\cal{D}}_0$. Choose an arbitrary half-plane $E\in{\cal{D}}''$ and assign
${\cal{D}}'':={\cal{D}}''\backslash\{E\}$ and $D^{(0)}:=E$. Consider a half-plane $\bar{E}\in {\cal{D}}'$ such
that the intersection $\bar{E}\cap l^{(0)}$ is a ray, where $l^{(0)}={\mathrm{bd}}\,E$. Let $\bar{h}$ be the
vector obtained by applying the procedure to the system ${\cal{D}}'\cup\{E\}$, the half- plane $D^{(0)}$, and the
ray on its border codirectional to the ray $\bar{E}\cap l^{(0)}$. If $\bar{h}\in E\cap
\bigcap\{D:D\in{\cal{D}}'\}$, then we assign ${\cal{D}}':={\cal{D}}'\cup\{E\}$ and $\bar{h}_0:=\bar{h}$. Choosing
an arbitrary half-plane $F\in{\cal{D}}''$, we set ${\cal{D}}'':={\cal{D}}''\backslash\{F\}$ and $D^{(0)}:=F$.
Consider a half-plane $\bar{F}\in {\cal{D}}'$ such that the intersection $\bar{F}\cap l^{(0)}$ is a ray, where
$l^{(0)}={\mathrm{bd}}\,F$. Let $\bar{h}$ be the vector obtained by applying the procedure to the system
${\cal{D}}'\cup\{F\}$, the half-plane $D^{(0)}$, and the ray on its border codirectional to the ray $\bar{F}\cap
l^{(0)}$. In the case of $\bar{h}\in F\cap \bigcap\{D:D\in{\cal{D}}'\}$, we set ${\cal{D}}':={\cal{D}}'\cup\{F\}$
and $\bar{h}_0:=\bar{h}$. We continue this process of constructing the system ${\cal{D}}'$ until
${\cal{D}}''\neq\varnothing$. After the process is finished, consider the subsystem ${\cal{J}}'$ of system
$(\ref{ks21})$ composed of all its inequalities whose sets of solutions are the half- planes of the system
${\cal{D}}'$. This subsystem is obviously coincides with an MCS of system $(\ref{ks21})$ containing the subsystem
${\cal{J}}_0$, and the vector $\bar{h}_0$ is a solution to the subsystem ${\cal{J}}'$.\clearpage

\section{ALGORITHM \\ FOR CONSTRUCTING A COMMITTEE}

Consider an inconsistent system of linear inequalities
\begin{equation}\label{ks51}
(c_j,h)>b_j, j=\overline{1,m}, c_j,h\in {\mathbf{R}}^2, m\geq 3,
\end{equation}
such that any system of two equations
$$(c_i,h)=b_i,\,(c_j,h)=b_j$$
is consistent, and any system of three equations $$(c_i,h)=b_i,\,(c_j,h)=b_j,\, (c_k,h)=b_k$$ with different $i$,
$j$, and $k$ is inconsistent.

{\bf Step 1. Finding marked MCSes.} Consider the system of half-planes ${\cal{D}}=\{D_j\}_{j=1}^m$, where $D_j$ is
the set of solutions to the $j$th inequality of system $(\ref{ks51})$, $j=\overline{1,m}$. Let $D\in {\cal{D}}$ be
an arbitrary half-plane and $r$ be a ray on its border. Assign $D^{(0)}:=D$. Applying the procedure described in
Section 2 to the half-plane $D^{(0)}$ and to the ray $r_0$ on its border codirectional to the ray $r$, we find a
marked MCS of system $(\ref{ks51})$. Applying the procedure to the half-plane $D^{(0)}$ and to the ray $r_0$ on
its border, counterdirectional to the ray $r$, we find, in the general case, another marked MCS of system
$(\ref{ks51})$. Repeat the same for any half-plane of the system ${\cal{D}}$. As a result, we obtain the set
$\{T_s\}_{s=1}^{n_0}$ of all (different) marked MCSes of system $(\ref{ks51})$ and the set $\{h'_s\}_{s=1}^{n_0}$
whose each element is a solution to the corresponding marked MCS. Step 1 is completed.

Consider the subsystem
\begin{equation}\label{ks52}
(c_j,h)>b_j, j\in \{k:(c_k,z)<b_k\},
\end{equation}
of system $(\ref{ks51})$, where the vector $z\in{\mathbf{R}}^2$ equates none
of the inequalities of system $(\ref{ks51})$.

{\bf Step 2. Constructing a minimal committee of subsystem $(\ref{ks52})$.} Find (see \cite{mazurov}) the index
sets for all MCSes of the system
\begin{equation}\label{ks53}
(c_j,h)>0, j\in \{k:(c_k,z)<b_k\},
\end{equation}
obtaining thereby the index sets for all MCSes of system $(\ref{ks52})$. Let ${\cal{J}}_1,\ldots,{\cal{J}}_{q_0}$
be the set of all (different) MCSes of system $(\ref{ks52})$, where $q_0$ is an odd integer. For each
$i=\overline{1,q_0}$, we define a vector $h^0_i$ as follows. If the subsystem ${\cal{J}}_i$ belongs to the marked
MCS $T_s$ for some $s=s(i)$, then we assign $h^0_i:=h'_s$. Moreover, if the subsystem ${\cal{J}}_i$ is contained
in two marked MCS’s, for example, in $s_1$st and $s_2$nd ones, then we take any of two elements $h'_{s_1}$ and
$h'_{s_2}$ as $h^0_i$. If the subsystem ${\cal{J}}_i$ is contained in none of the marked MCSes of system
(\ref{ks51}), then we find an $h^0_i$ to be equal to a solution of the found MCS.

Using \cite{mazurov}, we can show that the collection $K_0=\{h^0_i\}_{i=1}^{q_0}$ is a minimal committee of system
$(\ref{ks52})$. Consider the system ${\cal{J}}'={\cal{J}}'(K_0)$ consisting of all inequalities of system
$(\ref{ks51})$ such that each inequality of ${\cal{J}}'$ is not satisfied by $\frac{q_0+1}{2}$ elements of the
collection $K_0.$ Obviously, the system ${\cal{J}}'$ contains none of the inequalities of system $(\ref{ks52})$.
Moreover, using \cite{mazurov}, we can prove that each inequality of system $(\ref{ks51})$ not containing in
${\cal{J}}'$ is satisfied by at least $\frac{q_0+1}{2}$ elements of the collection $K_0$.

Construct the system of points ${\cal{D}}^{\ast}={\cal{A}}\cup{\cal{B}}$ corresponding to system $(\ref{ks51})$
with respect to the origin $z.$ Each inequality of system (\ref{ks52}) corresponds to a point of the system
${\cal{A}}$, and each inequality of system $(\ref{ks51})$ that does not belong to system $(\ref{ks52})$
corresponds to a point of the system ${\cal{B}}$. Consider the system of points ${\cal{B}}'\subset{\cal{B}}$
corresponding to the system ${\cal{J}}'$. To construct a committee of system $(\ref{ks51})$, it is sufficient to
find a committee of the system of points ${\cal{D}}^{\ast}$. Let $K^{\ast}_0=\{P^0_i\}_{i=1}^{q_0}$ where
$P^0_i=\{h:(h^0_i-z,h)>1\}$, $i=\overline{1,q_0}$. Assign $K^{\ast}:=K^{\ast}_0$ and
$\bar{{\cal{B}}}:={\cal{B}}'$. If ${\cal{B}}'=\varnothing$, then, obviously, $K^{\ast}=K^{\ast}_0$ is a committee
of the system of points ${\cal{D}}^{\ast}$ and the algorithm ends. Let ${\cal{B}}'\neq\varnothing$. Step 2 is
completed.

Below, by an angle with the vertex at $0$, we mean a closed convex cone bordered by two (possibly coinciding) rays
with a common vertex. These rays are referred to as sides of the angle. At the same time, the notion of angular
measure and adjacent angle are defined according to the standard terminology for the angles. At Step 3, new
elements will be added to the set of half- planes $K^{\ast}$.

{\bf Step 3. Consecutive search for members of the committee.} Draw a line through $0$ and a point $b\in
\bar{{\cal{B}}}$. Let ${\cal{A}}={\cal{A}}_1\cup {\cal{A}}_2$, where systems ${\cal{A}}_1={\cal{A}}_1(b)$ and
${\cal{A}}_2={\cal{A}}_2(b)$ are strictly separated by this line. Let ${\cal{B}}_1={\cal{B}}_1(b)$ be the system
consisting of all the points $b'\in \bar{{\cal{B}}}$ such that any line passing through $0$ and separating the
points $b$ and $b'$ strictly separates the systems ${\cal{A}}_1$ and ${\cal{A}}_2$.

By the construction of the set ${\cal{B}}_1$, there exists an angle with the vertex at $0$ and the angular measure
less than $\pi$ that contains this set. Let $\alpha$ be the minimal angle among all such angles. If ${\cal{B}}_1$
is a one-point set, then $\alpha$ is the zero angle. Extending the sides of the angle $\alpha$ to the lines,
consider the angles $\beta_1$ and $\beta_2$, which are adjacent to the angle $\alpha$. By construction of the set
${\cal{B}}_1$, any point of the system ${\cal{A}}$ belongs to one of the angles $\beta_1$ and $\beta_2$. For
definiteness, let ${\cal{A}}_1\subset\beta_2$ and ${\cal{A}}_2\subset\beta_1$. Let also
${\cal{B}}_{1k}={\cal{B}}\cap (\alpha\cup \beta_k)$. Consider the consistent subsystem $\bar{{\cal{J}}}_k$ of
system $(\ref{ks51})$ corresponding to the system of points ${\cal{A}}_k\cup{\cal{B}}_{1k}\cup\{0\}$, $k=1,2$. We
find a solution $\bar{h}_k$ to an MCS of system $(\ref{ks51})$ containing the subsystem $\bar{{\cal{J}}}_k$ and
assign $P_k=\{h:(\bar{h}_k-z,h)>1\}$, $k=1,2$.

If ${\cal{B}}_2={\cal{B}}\cap (P_1\cap P_2)\neq \varnothing$, then we construct two more half-planes. Let $\gamma$
be the minimal of the angles with the vertex at $0$ and the angular measure less than $\pi$ that contain the set
${\cal{B}}_2$. Extending the sides of the angle $\gamma$ to the lines, consider the angles $\beta'_1$ and
$\beta'_2$ adjacent to the angle $\gamma$. Let ${\cal{A}}_1\subset\beta'_2$ and ${\cal{A}}_2\subset\beta'_1$.
Consider the set ${\cal{B}}_{2k}={\cal{B}}\cap(\gamma\cup\beta'_k)$. Let $\bar{{\cal{J}}}_{k+2}$ be a consistent
subsystem of system $(\ref{ks51})$ corresponding to the system of points ${\cal{A}}_k\cup{\cal{B}}_{2k}\cup\{0\}$,
$k=1,2$. We find a solution $\bar{h}_{k+2}$ to an MCS of system $(\ref{ks51})$ containing the sub system
$\bar{{\cal{J}}}_{k+2}$ and assign $P_{k+2}=\{h:(\bar{h}_{k+2}-z,h)>1\}$, $k=1,2$.

We assign $K^{\ast}:=K^{\ast}\cup\{P_i\}$, where the collection $\{P_i\}$ contains two or four members. It is easy
to see that at least half of the half-planes of the union $\{P_i\}$ vote for each point of the system
${\cal{D}}^{\ast}$. We exclude from $\bar{{\cal{B}}}$ all the points for which the majority of members in
$\{P_i\}$ votes. If $\bar{{\cal{B}}}\neq\varnothing$, then go to Step 3. However, finally, case
$\bar{{\cal{B}}}=\varnothing$ arises, and then $K^{\ast}$ becomes a committee of the system ${\cal{D}}^{\ast}$.
The algorithm terminates. At the end, we have $K^{\ast}=\{Q_i\}_{i=1}^q$ for an odd integer $q$, where
$Q_i=\{h:(g_i-z,h)>1\}$, $g_i\in{\mathbf{R}}^2$, $i=\overline{1,q},$ then $\hat{K}=\{g_i\}_{i=1}^q$ is a committee
of system $(\ref{ks51})$.\clearpage

\section{AN EXISTENCE CRITERION \\ FOR A COMMITTEE OF THREE MEMBERS}

Consider an inconsistent system of linear inequalities
\begin{equation}\label{ks41}
(c_j,h)>b_j,\, c_j,h\in \mathbf{R}^2, j=1,\ldots,m,
\end{equation}
that has a committee. Let $|c_j|=1$, $j=\overline{1,m}$. We assume that system $(\ref{ks41})$ does not contain
identical inequalities. Let ${\cal{D}}$ be a system of half-planes which are the sets of solutions to the
corresponding inequalities of system (\ref{ks41}). Below, we apply the procedure described in Section 2.

{\bf Algorithm for finding a committee of three members.} Let $D^{(0)}\in{\cal{D}}$ be an arbitrary half-plane.
Consider a marked MCS $T_1$ of system $(\ref{ks41})$, its solution $\bar{h}_1$, and the determining pair of
inequalities $S_1$ that belongs to $T_1$, which are found by applying the procedure to the half-plane $D^{(0)}$
and to an arbitrary ray on its border. Let $E$ and $E'$ be the half-planes that are the sets of solutions to two
inequalities forming the subsystem $S_1$. Assign $D^{(0)}:=E$. Applying the procedure to the half-plane $D^{(0)}$
and to the ray on its border $l^{(0)}$ that is counterdirectional to the ray $E'\cap l^{(0)}$, we find a marked
MCS $T_2$ of system $(\ref{ks41})$ and its solution $\bar{h}_2$. Assign $D^{(0)}:=E'$. Consider the marked MCS
$T_3$ and its solution $\bar{h}_3$ that are found by applying the procedure to the half-plane $D^{(0)}$ and to the
ray on its border $l^{(0)}$ counterdirectional to the ray $E\cap l^{(0)}$. Let
$\bar{K}=\{\bar{h}_1,\bar{h}_2,\bar{h}_3\}$.

{\bf Proposition 4.1.} {\it System $(\ref{ks41})$ has a committee of three members if and only if the set
$\bar{K}$ is a committee of this system.}

{\bf Proof.} Let $K$ be a committee of system $(\ref{ks41})$ consisting of three members such that each its member
is a solution to an MCS of system $(\ref{ks41})$. From the proof of Proposition 2.4, it follows that the marked
MCSes $T_1$, $T_2$, and $T_3$ of system $(\ref{ks41})$ obtained by the algorithm are different. By Proposition
2.2, the marked MCS $T_s$ has a solution $h_s$ belonging to $K$ with multiplicity $1$, where $s=\overline{1,3}$.
Then, $K=\{h_1,h_2,h_3\}$. Since the vector $\bar{h}_s$ is a solution to the subsystem $T_s$, it follows that
$\bar{K}$ is a committee of system $(\ref{ks41})$. The proof of the second part of the proposition is obvious. The
proposition is proved.

Proving proposition 4.1, in fact, we also proved the
following proposition.

{\bf Proposition 4.2.}

(a) {\it Let $K$ be a committee of system $(\ref{ks41})$ consisting of three members such that each member is a
solution to an MCS of system $(\ref{ks41})$. Then, $K=\{h_1,h_2,h_3\}$, where $h_s$ is a solution to the subsystem
$T_s$, $s=\overline{1,3};$}

(b) {\it If system $(\ref{ks41})$ has a committee of three members and $K=\{h_1,h_2,h_3\}$, where $h_s$ is a
solution to the subsystem $T_s, s=\overline{1,3}$, then $K$ is a committee of system $(\ref{ks41})$.}

{\bf Theorem 4.1.} {\it System $(\ref{ks41})$ has a minimal committee of three members if and only if each of its
subsystems consisting of $5$ inequalities contains a consistent subsystem of four inequalities.}

{\bf Proof. Necessity.} Let $K$ be a minimal committee of system $(\ref{ks41})$ consisting of three members. Then,
$K$ is a committee of any its subsystem ${\cal{J}}'$ consisting of 5 inequalities. Therefore, as it was shown in
\cite{kobylkin1}, the subsystem ${\cal{J}}'$ contains a consistent subsystem of at least four inequalities.

{\bf Sufficiency.} Suppose that any subsystem of system $(\ref{ks41})$ consisting of five ine\-qua\-li\-ti\-es
contains a consistent subsystem of $4$ inequalities. Consider three marked MCSes $T_1$, $T_2$, and $T_3$ of system
$(\ref{ks41})$ and a set of points $\bar{K}=\{\bar{h}_1,\bar{h}_2,\bar{h}_3\}$, where $\bar{h}_i$ is the solution
to the subsystem $T_i$, $i=\overline{1,3}$, which are found by applying the construction described before
Proposition 4.1.

Prove that the $j$th inequality of system $(\ref{ks41})$ belongs to at least one of two subsystems $T_1$ and
$T_2$, where $j=1,\ldots,m$. Let $S_1$ and $S_2$ be the determining pairs of inequalities that are contained in
the subsystems $T_1$ and $T_2$, respectively. Uniting $S_1$ and $S_2$ in one subsystem and adding to it the $j$th
inequality, we obtain a system of five inequalities, which may contain identical ones. Consider a subsystem
${\cal{J}}'$ of this system that consists of all different inequalities. Since $T_1\neq T_2$, the set of solutions
$M_1$ to the pair of inequalities $S_1$ have no common points with the set of solutions $M_2$ to the pair of
inequalities $S_2$. Consequently, system ${\cal{J}}'$ is inconsistent and $|{\cal{J}}'|\geq 3$, where
$|{\cal{J}}'|$ is the cardinality of the system ${\cal{J}}'$. We show that this system contains a consistent
subsystem of cardinality $|{\cal{J}}'|-1$. For $|{\cal{J}}'|=5$, the existence of such a subsystem follows from
the hypothesis. Since the system ${\cal{J}}'$ has a committee, for $|{\cal{J}}'|=4$ and also for $|{\cal{J}}'|=3$,
the existence of such a subsystem follows from \cite{kobylkin1}. Then, from \cite{khachay} it follows that there
exists a committee $K'$ of system ${\cal{J}}'$ consisting of three elements. Since $M_1\cap M_2=\varnothing$, the
committee $K'$ has two different members, one of those satisfies the pair of inequalities $S_1$ and the other one
satisfies the pair of inequalities $S_2$. However, one of these two members, let us say the first one, satisfies
the $j$th inequality; hence, by Proposition 2.1, we find that the $j$th inequality belongs to $T_1$. Analogously,
it can be shown that the $j$th inequality belongs to at least one of two subsystems $T_2$ and $T_3$, and,
therefore, it can be shown that this inequality belongs to at least one of the subsystems $T_1$ and $T_3$.
Consequently, at least two members of $\bar{K}$ satisfy the $j$th inequality. Thus, $\bar{K}$ is a committee of
system $(\ref{ks41})$. The theorem is proved.

From \cite{khachay} it follows that, if the cardinality of MCS of system $(\ref{ks41})$ with the largest number of
inequalities is equal to $m-1$, then system $(\ref{ks41})$ has a committee of three members.

{\bf Corollary 4.1.} {\it Let the cardinality of MCS of system $(\ref{ks41})$ with the largest number of
inequalities be equal to $m-2$ and ${\cal{J}}_0$ be an MCS of cardinality $m-2$. Then, system $(\ref{ks41})$ has a
minimal committee of three members iff any its subsystem of $5$ inequalities containing both inequalities of the
system that do not belong to ${\cal{J}}_0$ contains a consistent subsystem of $4$ inequalities.}

Suppose that system $(\ref{ks41})$ satisfies the conditions imposed on it in Section 3. Consider the committee
$\hat{K}=\hat{K}(z)$ of system $(\ref{ks41})$ obtained by the algorithm for some $z\in {\mathbf{R}}^2$ that turns
into an equality none of its inequalities.

{\bf Proposition 4.3.} {\it If system $(\ref{ks41})$ has a committee of three members and the system
\begin{equation}\label{ks411}
(c_j,h)>b_j,\,j\in\{k:(c_k,z)<b_k\}
\end{equation}
is inconsistent, then $\hat{K}$ is a committee of system $(\ref{ks41})$ that consists of three members.}

{\bf Proof.} Let $K$ be a minimal committee of system $(\ref{ks41})$ consisting of three members such that each
its element is a solution to an MCS of system $(\ref{ks41})$. Then, $K$ is a minimal committee of system
$(\ref{ks411})$ in view of its inconsistency. Consider the set ${\cal{J}}_1,{\cal{J}}_2,{\cal{J}}_3$ of all
(different) MCSes of system $(\ref{ks411})$. Then, $K=\{h_1,h_2,h_3\}$, where $h_i$ is a solution to the subsystem
${\cal{J}}_i$, $i=\overline{1,3}$. It is easy to see that there are three marked MCSes of system $(\ref{ks41})$.
Therefore, at Step 1, the algorithm constructs the set $\{{\mathbf{T}}_s\}_{s=1}^{3}$ of the marked MCSes of
system $(\ref{ks41})$ and the set $\{h'_s\}_{s=1}^{3}$, each element of which is a solution to the corresponding
marked MCS. By Proposition 4.2 applied to the committee $K$, all members of this committee are solutions of the
marked MCSes of system $(\ref{ks41})$. Consequently, for any $i=1,2,3$, the subsystem ${\cal{J}}_i$ belongs to a
marked MCS of system $(\ref{ks41})$. At the same time, two subsystems ${\cal{J}}_{i_1}$ and ${\cal{J}}_{i_2}$,
$i_1\neq i_2$ cannot belong to the same marked MCS, since their sets of solutions are disjoint. Then, following
the algorithm at Step 2, we obtain a set $K_0=\{h'_1,h'_2,h'_3\}$ that is obviously a committee of system
$(\ref{ks41})$. Hence, the algorithm is completed at the end of Step 2 and $\hat{K}=K_0$. The proposition is
proved. \clearpage

\section{SYSTEMS ON THE BORDER\\
OF A CONVEX $m$-GON}

Consider an inconsistent system
\begin{equation}\label{ks31}
(c_j,h)>b_j,\, c_j,h\in \mathbf{R}^2, j=1,\ldots,m, m\geq 3,
\end{equation}
that has a committee. Let $|c_j|=1$, $c_i\neq c_j$, $i,j=\overline{1,m}$, $i\neq j$. Under the condition that the
lines $(c_j,h)=b_j,\,\,j=\overline{1,m},$ border a convex (possibly unbounded) polygon $M_0$ with $m$ sides, a
minimal committee of system $(\ref{ks31})$ will be constructed below. Transform this system by making the
substitution $h=h'+z$, where $z\in{\mathrm{int}}\,M_0$. The obtained system of linear inequalities has the form
\begin{equation}\label{ks311}
(c_j,h')>b'_j=b_j-(c_j,z),\,j=\overline{1,m}.
\end{equation}
Since the substitution $h'=h-z$ defines the parallel translation on the vector $-z$, the convex polygon bordered
by the lines $(c_j,h')=b'_j,\,\,j=\overline{1,m}$, has $m$ sides and contains $0$ in its interior. Any committee
of system (\ref{ks311}) can be easily transformed into a committee of system $(\ref{ks31})$. Thus, the
construction of a minimal committee of system $(\ref{ks31})$ will be made under the assumption that $0\in
{\mathrm{int}}\,M_0$.

Consider the subsystem
\begin{equation}\label{ks32}
(c_j,h)>b_j, j\in\{k:b_k>0\}
\end{equation}
of system $(\ref{ks31})$ that consists of all its inequalities with positive constant terms. Without loss of
generality, we may assume that this subsystem is composed of $k_1$ first inequalities of system $(\ref{ks31})$ for
some $k_1$, $1\leq k_1\leq m$. Below, a system ${\cal{J}}'$ will be constructed that contains system
$(\ref{ks31})$. At first, the system ${\cal{J}}'$ coincides with system $(\ref{ks31})$. In the process described
below, the new inequalities will be consecutively added to the system ${\cal{J}}'$.

We augment the system ${\cal{J}}'$ with the inequality $(c_{m+1},h)>b_{m+1}$, $c_{m+1}=c_1$, $b_{m+1}<0$, such
that the straight line $(c_{m+1},h)=b_{m+1}$ cuts off a convex figure from the polygon $M_0$. This figure is
either a triangle having a unique common vertex with $M_0$ or an unbounded region that contains no vertices of
$M_0$. The lines $(c_i,h)=b_i$, $i=\overline{1,m+1}$, border a convex polygon $M_1$ with $m+1$ sides that contains
$0$ in its interior. In the general case, for an arbitrary $j$, $2\leq j\leq k_1$, we add the inequality
$(c_{m+j},h)>b_{m+j}$, $c_{m+j}=c_j$, $b_{m+j}<0$, to the system ${\cal{J}}'$ under the condition that the line
$(c_{m+j},h)=b_{m+j}$ cuts off a convex figure from the polygon $M_{j-1}$. This figure can be either a triangle,
which has a unique common vertex with $M_{j-1}$, or an unbounded region, which contains no vertices of $M_{j-1}$.
The lines $(c_i,h)=b_i$, $i=\overline{1,m+j}$ bound a convex polygon $M_j$ with $m+j$ sides containing $0$ in its
interior. Adding in this way a new inequality to the system ${\cal{J}}'$ for $j=2$; then, for $j=3$, and so on
until $j=k_1$, we obtain a system ${\cal{J}}'$ of the form
\begin{equation}\label{ks33}
(c_j,h)>b_j,j=1,\ldots,m+k_1.
\end{equation}
Obviously, the polygon $M=M_{k_1}$ has $m+k_1$ sides and $0\in {\mathrm{int}}\,M$. Since system (\ref{ks31}) is
inconsistent, it follows that $k_1\geq 2$ and, consequently, the polygon $M$ is bounded. Below, two sides of the
polygon $M$ are called {\it neighbor}, if they are incident to a common vertex of $M$.

{\bf Algorithm for constructing a minimal committee.} Let ${\cal{D}}=\{D_j\}_{j=1}^{m+k_1}$ be a system of
half-planes, each one being the set of solutions to the corresponding inequality of system $(\ref{ks33})$.
Consider all possible pairs of neighbor sides of the polygon $M$ such that one of the sides of each pair lies on
the border of some half-plane of system ${\cal{D}}$ containing 0 and the other side of this pair lies on the
border of some half-plane from ${\cal{D}}$ that does not contain $0$. Denote by $p$ the number of pairs of the
neighboring sides of the polygon $M$ satisfying this condition. For each $i$, $1\leq i\leq p$, consider a
half-plane $D^1_i\in{\cal{D}}$, $0\notin D^1_i$ whose border $l^1_i$ contains one of the sides of the $i$th pair,
and a half-plane $D^2_i$, $0\in D^2_i$ whose border contains the other side of the $i$th pair. Let $r_i=D^2_i\cap
l^1_i$. The intersection $\bigcap\{D\cap l^1_i: D\in{\cal{D}}, D\cap r_i \mbox{is a ray}\}$ coincides with a ray
$E_i\cap l^1_i$, $E_i\in{\cal{D}}$ whose vertex will be denoted by $h_i$. Take an element $\bar{h}_i\in D^1_i\cap
E_i$ in a small neighborhood of the point $h_i$ such that $\bar{h}_i\in\bigcap\{D\in{\cal{D}}:h_i\in D\}$. The
collection $L=\{h_i\}_{i=1}^p$ contains repeating terms. Let $L'=\{h_{i_k}\}_{k=1}^q$ be a set of all (different)
elements of the collection $L$, $q<p$, $\bar{K}=\{\bar{h}_{i_k}\}_{k=1}^q.$
\medskip

{\bf Theorem 5.1.} {\it The collection $\bar{K}$ is a minimal committee of system $(\ref{ks31})$. The number $q$
of its members is equal to $p-q_0$, where $q_0$ is the number of members of the minimal committee of system
$(\ref{ks32})$.}

See Appendix for the proof.

Suppose that system $(\ref{ks31})$ satisfies the conditions imposed on this system in Section 3. Consider the
committee $\hat{K}=\hat{K}(z)$ of system $(\ref{ks31})$ found by the algorithm for some $z\in {\mathbf{R}}^2$ that
turns into an equality none of its inequalities.

{\bf Corollary 5.1.1.} {\it If the origin $z$ lies inside a convex $m$-gon bounded by straight lines
$(c_j,h)=b_j$, $j=\overline{1,m}$, then $\hat{K}$ is a minimal committee of system $(\ref{ks31})$.}

\section{CONCLUSIONS}

In the present work, we propose the representation
of an arbitrary system of strict linear inequalities in Rn
as a system of points obtained by the so-called polarity.
An algorithm for constructing a committee of a plane
system of linear inequalities is given based on this representation.
The solution of two problems on a minimal
committee of plane systems is proposed. In particular,
the solution of the problem on a minimal committee of
three members is given. The found solutions to these
problems can be obtained by the proposed algorithm.
\bigskip

\begin{center}
ACKNOWLEDGMENTS
\end{center}

This work was supported by the Russian Foundation for Basic Research, project nos. NSh-792.2003.1, 04-01-00108,
and 01-01-00563.

\renewcommand{\refname}

{\small Kobylkin K.S. Born in 1977. Graduated from Ural State University in 2000. At the present time Kobylkin
K.S. works at Institute of Mathematics and Mechanics of the Ural Division of the Russian Academy of Sciences.
Earned his Ph.D. in 2005. Scientific interests: pattern recognition, mathematical economics. Author of four
publications. Member of the Russian Association for Pattern Recognition and Image Analysis and IAPR.} \clearpage

\section{Appendix: proof of the Theorem 5.1}

Using \cite{mazurov}, we can prove the following lemma.

{\bf Lemma 5.1.} {\it Let ${\cal{J}}_1,\ldots,{\cal{J}}_{q_1}$ be the set of all (different) MCSes of system
$(\ref{ks32})$. Then

(a) $q_1=q_0$, and any minimal committee $K$ of system $(\ref{ks32})$ is given by $K=\{h^s\}_{s=1}^{q_0}$, where
$h^s$ is a solution to the subsystem ${\cal{J}}_s$, $s=\overline{1,q_0}$. Any inequality of system $(\ref{ks32})$
is satisfied by exactly $\frac{q_0+1}{2}$ members of the committee $K$;

(b) the set of solutions to any subsystem ${\cal{J}}_s,$ $1\leq s\leq q_0,$ coincides with a cone $D'\cap D''$,
where $D'=D'(s)$ and $D''=D''(s)$ are half-planes of the system ${\cal{D}}$ that do not contain $0$.}

{\bf Lemma 5.2.} {\it The set of solutions to any MCS of system $(\ref{ks33})$ coincides with the intersection
$D^1_i\cap D^2_i\cap D^1_j\cap D^2_j$ for certain different $i$ and $j$, $1\leq i,j\leq p$. Conversely, if
$N_{ij}=D^1_i\cap D^2_i\cap D^1_j\cap D^2_j\neq\varnothing$, then $N_{ij}$ coincides with the set of solutions to
some MCS of system $(\ref{ks33})$.}

{\bf Proof of Theorem 5.1.} Consider the set ${\cal{J}}_1,\ldots,{\cal{J}}_{q_0}$ of all (different) MCSes of
system $(\ref{ks32})$. Lemma 5.1 implies that the set of solutions to the subsystem ${\cal{J}}_s$,
$s=\overline{1,q_0},$ coincides with a cone bounded by two half-planes of the system ${\cal{D}}$ not containing
$0$. Let $K_0=\{g_s\}_{s=1}^{q_0}$ where $g_s$ is the vertex of the cone of solutions of subsystem ${\cal{J}}_s$,
$s=\overline{1,q_0}$. We prove that, if $h_i=h_j=h_0$, $h_0\in{\mathbf{R}}^2$, $1\leq i,j\leq p$, $i\neq j$, then
$h_0=g_{s}$ for some $s$, $1\leq s\leq q_0$. By the construction of $h_i$ and $h_j$, the point $h_0$ is the common
vertex of the rays $E_i\cap l^1_i$ and $E_j\cap l^1_j$ that coincide with the intersections $\bigcap\{D\cap
l^1_i:D\in{\cal{D}},D\cap r_i \mbox{is a ray}\}$ and $\bigcap\{D\cap l^1_j:D\in{\cal{D}},D\cap r_j \mbox{is a
ray}\}$, respectively. Since the polygon $M$ has $m+k_1$ sides, it follows that $h_0\notin{\mathrm{bd}}\,D$ for
any half-plane $D\in{\cal{D}}$ different from $D^1_i$ and $D^1_j$. Therefore, $E_i=D^1_j$ and $E_j=D^1_i$. Then,
by Corollary 2.3.1, the half-planes $D^1_i$ and $D^1_j$ are the sets of solutions to the inequalities that form a
determining pair for a marked MCS of system $(\ref{ks33})$.

Let $D\in{\cal{D}}$ be an arbitrary half-plane that does not contain $0$. If $D^1_i\cap D^1_j\cap
D\neq\varnothing$, then $h_0\in D$. Since $M\subseteq ({\mathbf{R}}^2\backslash D^1_i)\cap
({\mathbf{R}}^2\backslash D^1_j)\cap ({\mathbf{R}}^2\backslash D)$, one of the three half-planes $D^1_i$, $D^1_j$,
and $D$ contains the intersection of the other two. Consequently, $D^1_i\cap D^1_j\subset D$. Then, the
intersection $D^1_i\cap D^1_j$ coincides with the cone of solutions to some MCS of system $(\ref{ks32})$, i.e.,
$h_0=g_s$ for some $s$, $1\leq s\leq q_0$.

Let $h_0$ be the vertex of the cone of solutions to an arbitrary MCS of system $(\ref{ks32})$. This cone coincides
with the intersection $D^1\cap D^2$, where $D^1$ and $D^2$ are two half-planes of the system ${\cal{D}}$ that do
not contain $0$. We show that $h_0=h_i=h_j$ for certain different $i$ and $j$, $1\leq i,j\leq p$. Consider the
side $u_1$ of the polygon $M$ that lies on the border $l^1$ of the half-plane $D^1$ and the endpoint $h'$ of this
side such that $u_1$ lies on the interval $[h_0,h']$. Take a half-plane $D'\in{\cal{D}}$ such that its border $l'$
passes through the point $h'$. We prove that $0\in D'$. To do this, we show that the line $l'$ is either parallel
to the line $l^2={\mathrm{bd}}\,D^2$ or intersects this line at a point that belongs to the ray $D^1\cap l^2$.
Consider the half-plane $D^0\in {\cal{D}}$ containing $0$ such that the line $l^0={\mathrm{bd}}\,D^0$ is parallel
to the line $l^2$. If $l'=l^0$, then $D'=D^0$. Let $l'\neq l^0$. Consider the triangle $T$ bounded by three lines
$l'$, $l^1$ and $l^2$. On the one hand, if $T$ and $M$ lie to the same side of the line $l^1$, then $M\subseteq
T$. On the other hand, the line $l^0$ contains a side of the polygon $M$. Since the vertices of the triangle $T$
belong to the half-plane $D^0$, $T$ and $M$ belong to the opposite sides with respect to the line $l^1$.
Consequently, the intersection point of the lines $l'$ and $l^2$ lies on the ray $D^1\cap l^2$. Since $D^1\cap
D^2$ is the cone of solutions to an MCS of system $(\ref{ks32})$, we have either $D^1\cap D^2\subset D$ or
$D^1\cap D^2\cap D=\varnothing$ for any half-plane $D\in {\cal{D}}$, $0\notin D$. Consequently, $0\in D'$. Since
the sides of the polygon $M$ lying on the lines $l^1$ and $l'$ are neighboring, $D^1=D^1_i$ and $D'=D^2_i$ for
some $i$, $1\leq i\leq p$. By construction, the point $h_i$ is the vertex of the ray $E_i\cap l^1=\bigcap\{D\cap
l^1:D\in{\cal{D}},D\cap D'\cap l^1 \mbox{is a ray}\}$. Since $u_1\subset{\mathrm{cl}}\,D'\cap l^1$, we have
$h_0\in D'$ and, consequently, $D^2\cap l^1\subset D'\cap l^1$. Therefore, $E_i\cap l^1\subseteq D^2\cap l^1$.
Then, the half-plane $E_i$ is disjoint with the side $u_1$, i.e., $0\notin E_i$. Since $E_i\cap D^1\cap
D^2\neq\varnothing$, we have $D^1\cap D^2\subset E_i$. Then, $E_i\cap l^1=D^2\cap l^1$ and, consequently,
$E_i=D^2$ and $h_0=h_i$.

Consider the side $u_2$ of the polygon $M$ lying on the border of the half-plane $D^2$. Let $h''$ be an endpoint
of the side $u_2$ such that $u_2$ lies on the interval $[h_0,h'']$. By similar arguments, we can show that the
half-plane $D''\in {\cal{D}}$ whose border passes through the point $h''$ contains $0$. Then, $D^2=D^1_j$ and
$D''=D^2_j$ for some $j$, $i\neq j$, $1\leq j\leq p$. It can be similarly shown that $h_0=h_j$ and $E_j=D^1$.
Since $l^1_i\cap l^1_j\cap l^1_k=\varnothing$ for any $k\neq i,j$, it follows that $h_0\neq h_k$. Therefore, the
elements of the collection $K_0$ have multiplicity $2$ in $L$ and the elements of the collection $L'$ that do not
belong to $K_0$ have multiplicity $1$. Then, $q=p-q_0$. Let for definiteness, $L'=\{h_i\}_{i=1}^{q}$,
$\bar{K}=\{\bar{h}_i\}_{i=1}^q$ and $h_{q+s}=h_s=g_s$, $s=\overline{1,q_0}$.

Consider the system of points ${\cal{D}}^{\ast}={\cal{A}}\cup{\cal{B}}$ that corresponds to system $(\ref{ks33}).$
Denote by ${\cal{B}}_0\subset{\cal{B}}$ the subset of points corresponding to the inequalities of this system with
the numbers $m+1,\ldots,m+k_1.$ Let $M^{\ast}={\mathrm{conv}}\,{\cal{D}}^{\ast}.$ Since $0\in{\mathrm{int}}\,M,$
it follows that $M^{\ast}$ is the polar of $M,$ $0\in {\mathrm{int}}\,M^{\ast}.$ Since the polygon $M$ has $m+k_1$
sides, each point other than $0$ of the system ${\cal{D}}^{\ast}$ is a vertex of $M^{\ast}.$ The border of
$M^{\ast}$ is a closed broken line. Denote it by $\Gamma.$ Let $P_i=\{h:(h_i,h)\geq 1\}$ and
$\delta_i=P_i\cap\Gamma$, $i=\overline{1,p}.$ Consider the points ${\mathbf{a}}_i\in{\cal{A}}$ and
${\mathbf{b}}_i\in{\cal{B}}$ corresponding to the half-planes $D^1_i$ and $D^2_i,$ respectively,
$i=\overline{1,p}.$ Since the sides of $M$ lying on the borders of the half-planes $D^1_i$ and $D^2_i$ are
neighboring, ${\mathbf{a}}_i$ and ${\mathbf{b}}_i$ are neighboring vertices of $M^{\ast}.$ Obviously, if
$E_i=D^2_i,$ then the broken line $\delta_i$ coincides with the edge $[{\mathbf{a}}_i,{\mathbf{b}}_i]$ of the
polygon $M^{\ast}.$ Let us find the shape of the broken line $\delta_i$ in the case where $E_i\neq D^2_i.$ The
line that passes through $0$ and ${\mathbf{a}}_i$ divides the broken line $\Gamma$ into two parts, $\Gamma^1_i$
and $\Gamma^2_i.$ Let ${\mathbf{b}}_i\in\Gamma^2_i$ for any $i=\overline{1,p}.$ Obviously, the parts $\Gamma^1_i$
and $\Gamma^2_i$ have two common endpoints: one of them coincides with ${\mathbf{a}}_i$, and the other one
coincides with a point ${\mathbf{b}}'_i\in{\cal{B}}_0$. By construction, the element $h_i$ is the vertex of the
ray $E_i\cap l^1_i=\bigcap\{D\cap l^1_i:D\in{\cal{D}},D\cap r_i\; \mbox{is a ray}\}.$ It is easy to see that, for
any half-plane $D\in{\cal{D}},$ the intersection $D\cap r_i$ is a ray if and only if the point of the system
${\cal{D}}^{\ast}$ that corresponds to this half-plane belongs to one of the sets $\Gamma^1_i\cap{\cal{A}}$ or
$\Gamma^2_i\cap{\cal{B}}.$ Since $E_i\neq D^2_i,$ we have $0\notin E_i.$ Hence, the line $(h_i,h)=1$ passes
through two different points lying on $\Gamma^1_i:$ one of them coincides with ${\mathbf{a}}_i,$ and the other
coincides with a point ${\mathbf{a}}'_i\in {\cal{A}},$ corresponding to the half-plane $E_i.$ Since
$\Gamma^1_i\cap{\cal{A}}\subset P_i$ by the construction of element $h_i,$ we have
$\Gamma^1_i\cap{\cal{A}}=\delta_i\cap{\cal{A}}.$ Consequently, $\delta_i,$ being a part of the broken line
$\Gamma^1_i,$ connects the points ${\mathbf{a}}_i$ and ${\mathbf{a}}'_i.$

Calculate now the number $u(j)$ of members of the collection $\bar{K}$ satisfying the $j$th inequality of system
$(\ref{ks33}),$ where $j=1,\ldots,m+k_1.$ This calculation will be done in two steps. First, calculate the number
$v(j)$ of members of the collection $\bar{L}=\{\bar{h}_i\}_{i=1}^p$ satisfying this inequality. Then, calculate
the number $w(j)$ of members of the collection $\bar{K}_0=\{\bar{h}_{q+s}\}_{s=1}^{q_0}$ satisfying the same
inequality. We obtain $u(j)=v(j)-w(j)$. For any $i$, $1\leq i\leq p,$ we have $h_i\notin {\mathrm{bd}}\,D$ for any
half-plane $D\in{\cal{D}}$ that differs from $D^1_i$ and $E_i$, $i=\overline{1,p}$. Thus, by construction of the
element $\bar{h}_i,$ we have $\{D\in{\cal{D}}:\bar{h}_i\in D\}=\{D\in{\cal{D}}:h_i\in D\}\cup\{D^1_i,E_i\}.$ Then,
the number of elements of the collection $\bar{L}$ satisfying the $j$th inequality coincides with the members of
terms of the collection $L,$ which either satisfy the $j$th inequality or turn it into an equality.

Calculate the number $v(j)$ in the case where the $j$th inequality cor\-res\-ponds to a point
${\mathbf{a}}\in{\cal{A}}$. Consider the family of broken lines $\Delta=\{\delta_i\}_{i=1}^{p}$. The number $v(j)$
is equal to the number of the broken lines in the family $\Delta$ containing the point ${\mathbf{a}}$. Consider on
the boundary of $M^{\ast}$ broken lines $\Gamma$ that have their endpoints in ${\cal{A}}$ and contain no points of
${\cal{B}}$. Among them we choose $\Gamma_1,\ldots,\Gamma_{p_0}$ that are maximal by inclusion. Obviously,
$p=2p_0$. It is easy to see that the endpoints of any broken line $\Gamma_t$, $1\leq t\leq p_0$, are two points
${\mathbf{a}}_k$ and ${\mathbf{a}}_{n}$ (perhaps coinciding) of ${\cal{A}}$ for two different $k$ and $n$, $1\leq
k,n\leq p$. We show that ${\mathbf{a}}\in\Gamma_t$ if and only if ${\mathbf{a}}\in\delta_k\cap\delta_n$. If
${\mathbf{a}}\in\Gamma_t$, then ${\mathbf{a}}\in\delta_k$, since $\Gamma_t\subseteq\Gamma^1_k$ and
$\Gamma^1_k\cap{\cal{A}}=\delta_k\cap{\cal{A}}$. Analogously, it can be shown that ${\mathbf{a}}\in\delta_n$.
Conversely, if ${\mathbf{a}}\in\delta_k\cap\delta_n$, then ${\mathbf{a}}\in\Gamma^1_k\cap\Gamma^1_n$. If
${\mathbf{a}}_k\neq {\mathbf{a}}_n$, then $\Gamma^1_k\cap\Gamma^1_n=\Gamma_t$, and, in the case
${\mathbf{a}}_k={\mathbf{a}}_n$, we have $\Gamma^1_k\cap\Gamma^1_n\cap{\cal{A}}=\{{\mathbf{a}}_k\}=\Gamma_t$,
hence ${\mathbf{a}}\in\Gamma_t$.

Since $\Gamma_t\cap{\cal{B}}_0=\varnothing$, the part of the broken line $\Gamma^2_k$ connecting the points
${\mathbf{b}}'_k$ and ${\mathbf{b}}'_n$ contains no points of the system ${\cal{A}}$. Therefore, if
${\mathbf{a}}\notin\Gamma_t$, then ${\mathbf{a}}$ belongs to just one of two broken lines $\delta_k$ and
$\delta_n$. Moreover, obviously, the point ${\mathbf{a}}$ belongs to just one broken line of the family
$\{\Gamma_t\}_{t=1}^{p_0}$. Summing up over all the broken lines of this family, we obtain that ${\mathbf{a}}$
belongs to exactly $p_0+1$ broken lines of the family $\Delta$, i.e., $v(j)=p_0+1$. Calculate now the number
$w(j)$. Since $h_{q+s}=g_s$, $s=\overline{1,q_0}$, by construction of the element $\bar{h}_{q+s}$ and by Lemma
5.1, the set $\bar{K}_0$ is a minimal committee of system $(\ref{ks32})$. At the same time, since the $j$th
inequality belongs to system $(\ref{ks32})$, exactly $\frac{q_0+1}{2}$ elements of the committee $\bar{K}_0$
satisfy this inequality; hence, $w(j)=\frac{q_0+1}{2}$. Then, $u(j)=p_0+1-\frac{q_0+1}{2}=\frac{p-q_0+1}{2}$.

Calculate the number $v(j)$ in the case where a point ${\mathbf{b}}\in{\cal{B}}$ corresponds to the $j$th
inequality. Consider the family of broken lines $\Delta'=\{\delta'_i\}_{i=1}^p,$ where
$\delta'_i=\{h\in\Gamma:(h_i,h)\leq 1\}$, $i=\overline{1,p}$. Then, the number $v(j)$ is equal to the number of
broken lines of the family $\Delta'$ containing the point ${\mathbf{b}}$. Let $\Gamma({\mathbf{b}})\subset \Gamma$
be the minimal by inclusion part (containing the point ${\mathbf{b}}$) of the broken line $\Gamma$ with the
endpoints at the points of system ${\cal{A}}$. Consider the subcase where there exists a point
${\mathbf{b}}_0\in{\cal{B}}$ that belongs to the intersection $\Gamma({\mathbf{b}})\cap{\cal{B}}_0$. Denote by
${\mathbf{a}}_0$ a point of the system ${\cal{A}}$ that lies on the line passing through $0$ and ${\mathbf{b}}_0$.
It is easy to see that if ${\mathbf{a}}_0\in\delta_i$, then ${\mathbf{b}}_0\in\delta'_i$, $i=\overline{1,p}$.
Moreover, by construction of the broken line $\Gamma({\mathbf{b}})$, one of two inclusions takes place: either
$\Gamma({\mathbf{b}})\subseteq\delta_i$ or $\Gamma({\mathbf{b}})\subseteq\delta'_i$ for any $i=\overline{1,p}$.
Since ${\mathbf{b}}_0\in\Gamma({\mathbf{b}})$, it follows that ${\mathbf{b}}\in\delta'_i$ if and only if
${\mathbf{b}}_0\in\delta'_i$, $i=\overline{1,p}$. Let point ${\mathbf{a}}_0$ correspond to the $j_0$th inequality
of system $(\ref{ks32})$ for some $j_0$, $1\leq j_0\leq k_1$. Then, $v(j)\geq v(j_0)=p_0+1$. Calculate the number
$w(j)$. Consider an arbitrary point ${\mathbf{a}}'\in{\cal{A}}$ and an arbitrary broken line $\delta_{q+s}$,
$1\leq s\leq q_0$, such that ${\mathbf{a}}'\notin \delta_{q+s}$. Let ${\mathbf{b}}'$ be a point of the system
${\cal{B}}_0$ that lies on the line passing through $0$ and ${\mathbf{a}}'$. We show that
${\mathbf{b}}'\in\delta_{q+s}$. Since $h_{q+s}=h_s=l^1_s\cap l^1_{q+s}$, the endpoints of the broken line
$\delta_{q+s}$ are two different points ${\mathbf{a}}_s$ and ${\mathbf{a}}_{q+s}$. Then,
$\delta_{q+s}=\Gamma^1_s\cap\Gamma^1_{q+s}.$ Since
$\Gamma^1_s\cap{\cal{A}}=\Gamma^1_{q+s}\cap{\cal{A}}=\delta_{q+s}\cap{\cal{A}}$, we have
${\mathbf{a}}'\in\Gamma^2_s\cap\Gamma^2_{q+s}$, whence ${\mathbf{b}}'\in\Gamma^1_s\cap\Gamma^1_{q+s}$, i.e.,
${\mathbf{b}}'\in\delta_{q+s}$. Thus, the number of broken lines of the family $\{\delta_{q+s}\}_{s=1}^{q_0}$
containing the point ${\mathbf{a}}_0$ is equal to the number of broken lines of the family
$\{\delta'_{q+s}\}_{s=1}^{q_0}$ containing the point ${\mathbf{b}}_0$. Since the first of these numbers equals
$w(j_0)$ and the second one equals $w(j)$, by the above-proved, we have $w(j)=w(j_0)=\frac{q_0+1}{2}$. Hence,
$u(j)\geq u(j_0)=\frac{p-q_0+1}{2}$.

Consider now the subcase where $\Gamma({\mathbf{b}})\cap{\cal{B}}_0=\varnothing$ and calculate the number $v(j)$.
Take an arbitrary broken line $\Gamma_t$, $1\leq t\leq p_0$. Let the points ${\mathbf{a}}_k$ and
${\mathbf{a}}_{n}$ of system ${\cal{A}}$ be the endpoints of this broken line, $k\neq n$, $1\leq k,n\leq p$. We
show that ${\mathbf{b}}$ belongs to just one of the two broken lines $\delta'_k$ and $\delta'_n$. If two endpoints
of each of the two broken lines $\delta_k$ and $\delta_n$ are the points of the system ${\cal{A}}$, then
$\Gamma_t=\delta_k\cap\delta_n$. Consequently, if ${\mathbf{b}}\in\delta_k\cap\delta_n$, then one of the two
broken lines $\delta_k$ or $\delta_n$ coincides with the edge of the polygon $M^{\ast}$ and the point
${\mathbf{b}}$ is a vertex incident to this edge. Therefore, the point ${\mathbf{b}}$ belongs to at least one
broken line $\delta'_k$ or $\delta'_n$. On the other hand, since $\Gamma^1_k\cap{\cal{A}}=\delta_k\cap{\cal{A}}$,
$\Gamma^1_n\cap{\cal{A}}=\delta_n\cap{\cal{A}}$, and $\Gamma_t\cap{\cal{B}}_0=\varnothing$, we have
$\Gamma({\mathbf{b}}'_k)=\delta'_k\cap\delta'_n$. Since $\Gamma({\mathbf{b}})\cap{\cal{B}}_0=\varnothing,$ it
follows that ${\mathbf{b}}\notin\delta'_k\cap\delta'_n$. Summing up over all the broken lines of the family
$\{\Gamma_t\}_{t=1}^{p_0}$, we find that the point ${\mathbf{b}}$ belongs to exactly $p_0$ broken lines of the
family $\Delta'$, whence $v(j)=p_0$.

Let us calculate the number $w(j)$. We show that the broken line $\Gamma({\mathbf{b}})$ lies in the intersection
of $\frac{q_0+1}{2}$ broken lines of the family $\{\delta_{q+s}\}_{s=1}^{q_0}$. Indeed, consider an arbitrary
broken line $\delta_{q+s}$ containing one of the endpoints of the part $\Gamma({\mathbf{b}})$ that coincides with
a point ${\mathbf{a}}_0\in{\cal{A}}$. If ${\mathbf{a}}_0$ is neither endpoint of the broken line $\delta_{q+s}$,
then $\Gamma({\mathbf{b}})\subseteq \delta_{q+s}$ by construction of $\Gamma({\mathbf{b}})$. In the case where
${\mathbf{a}}_0$ is an endpoint of the broken line ${\mathbf{a}}_0$, we have either
${\mathbf{a}}_0={\mathbf{a}}_s$ or ${\mathbf{a}}_0={\mathbf{a}}_{q+s}$. Let, for definiteness,
${\mathbf{a}}_0={\mathbf{a}}_s$. If $\Gamma({\mathbf{b}})\subset\delta'_{q+s}$, then
${\mathbf{b}}'_{q+s}\in\Gamma({\mathbf{b}})\cap{\cal{B}}_0$ since
$\Gamma^1_{q+s}\cap{\cal{A}}=\delta_{q+s}\cap{\cal{A}};$ however, this is impossible. Thus,
$\Gamma({\mathbf{b}})\subseteq\delta_{q+s}$. Consequently, the point ${\mathbf{b}}$ belongs to exactly
$\frac{q_0+1}{2}$ broken lines of the family $\{\delta_{q+s}\}_{s=1}^{q_0}.$ Then, the point ${\mathbf{b}}$
belongs to exactly $\frac{q_0-1}{2}$ broken lines of the family $\{\delta'_{q+s}\}_{s=1}^{q_0}$. Consequently,
$w(j)=\frac{q_0-1}{2}$ i.e., $u(j)=\frac{p-q_0+1}{2}$. Thus, $\bar{K}$ is a committee of system $(\ref{ks33})$.

Let us prove that $\bar{K}$ is a minimal committee. Consider an arbitrary committee
$K_1=\{\bar{f}_k\}_{k=1}^{q_1}$ of system $(\ref{ks33})$ such that all its elements are so\-lu\-ti\-ons to some
MCSes of system $(\ref{ks33})$. We show that $q_1\geq p-q_0$. The set of solutions to the subsystem ${\cal{J}}_s$,
$1\leq s\leq q_0,$ coincides with the set of solutions to a pair of its inequalities, which is, as has been proved
above, the determining pair of a marked MCS $T_s$ of system $(\ref{ks33})$. Since $K_1$ is a committee of system
$(\ref{ks33})$, the collection $K^{\ast}_1=\{Q_k\}_{k=1}^{q_1}$ is a committee of the system ${\cal{D}}^{\ast}$.
By Proposition 2.2 applied to the committee $K_1$, the subsystem $T_s$ has a solution entering the committee $K_1$
with a positive multiplicity. Therefore, there exists a member $Q^s\in K^{\ast}_1$ that votes for the subsystem
$({\cal{A}}\cap\delta_s)\cup({\cal{B}}\cap\delta'_s)$ of the system of points ${\cal{D}}^{\ast}$ corresponding to
the marked MCS $T_s$. Since, by construction, the element $\bar{h}_s$ satisfies the subsystem $T_s$ without loss
of generality, we may assume that $Q^s=Q_s=\bar{P}_s$, where $\bar{P}_s=\{h:(\bar{h}_s,h)>1\}$,
$s=\overline{1,q_0}$. Moreover, for all $k=\overline{1,q_1}$ such that $\{j:1\leq j\leq
m+k_1,(c_j,\bar{f}_k)>b_j\}=\{j:1\leq j\leq m+k_1,(c_j,\bar{h}_i)>b_j\}$ for some $i$, $1\leq i\leq p$, we assume
without loss of generality that $Q_k=\bar{P}_i=\{h:(\bar{h}_i,h)>1\}$. Consider the sets $I_1=\{i:1\leq i\leq
p,\Gamma({\mathbf{b}}_i)\cap{\cal{B}}_0\neq\varnothing\}$ and $I_2=\{i:1\leq i\leq
p,\Gamma({\mathbf{b}}_i)\cap{\cal{B}}_0=\varnothing\}$. We show that, for any $i\in I_2$, there exists a
half-plane $Q\in K'_1=\{Q_k\}_{k=q_0+1}^{q_1}$ that votes for the points ${\mathbf{a}}_i$ and ${\mathbf{b}}_i$.
Since $K^{\ast}_1$ is a committee of the system of points ${\cal{D}}^{\ast}$, there exists a half-plane $Q\in
K^{\ast}_1$ which votes for these two points. Moreover, for any $s=\overline{1,q_0}$, the half-plane $Q_s$ votes
for two points ${\mathbf{a}}_s$ and ${\mathbf{b}}_s$, and also for two points ${\mathbf{a}}_{q+s}$ and
${\mathbf{b}}_{q+s}$; moreover, ${\mathbf{b}}'_{q+s}\in\Gamma({\mathbf{b}}_{s})\cap{\cal{B}}_0$ and
${\mathbf{b}}'_{s}\in\Gamma({\mathbf{b}}_{q+s})\cap{\cal{B}}_0$. Consequently, the line ${\mathrm{bd}}\,Q_s$
intersects $\Gamma$ at two points lying on the edges $[{\mathbf{a}}_s,{\mathbf{b}}_s]$ and
$[{\mathbf{a}}_{q+s},{\mathbf{b}}_{q+s}]$. Then, $Q_s$ cannot vote for both points ${\mathbf{a}}_i$ and
${\mathbf{b}}_i$. Therefore, $Q\in K'_1$.

Lemma 5.2 implies that $Q=\bar{P}_i$ for $\Gamma^1_i\cap{\cal{A}}=\{{\mathbf{a}}_i\}$. Let
$\Gamma^1_i\cap{\cal{A}}\neq\{{\mathbf{a}}_i\}$. We show that the entire broken line $\delta_i$ lies in a broken
line $\delta_{s_0}$, $1\leq s_0\leq q_0$, which has a common endpoint ${\mathbf{a}}'_i$ with $\delta_i$. Then, we
show that $Q\neq\bar{P}_i$ if and only if $Q$ votes against two points of the system ${\cal{A}}$ that are the
endpoints of the broken line $\delta_{s_0}$. Indeed, since the part of the broken line $\Gamma^1_i$ connecting the
points ${\mathbf{a}}'_i$ and ${\mathbf{b}}'_i$ contains no points of the system ${\cal{A}}$ other than
${\mathbf{a}}'_i$, we have ${\mathbf{b}}'_i\in\Gamma({\mathbf{b}}_0)\cap{\cal{B}}_0$, where ${\mathbf{b}}_0$ is a
point neighboring with ${\mathbf{a}}'_i$ of the system ${\cal{B}}$ lying on this part. As has been proved above,
$\bar{K}^{\ast}_0=\{Q_s\}_{s=1}^{q_0}$ is a committee of the system of points
${\cal{A}}\cup\{{\mathbf{b}}\in{\cal{B}}:\Gamma({\mathbf{b}})\cap{\cal{B}}_0\neq\varnothing\}\cup\{0\}$. Hence,
there exists a half-plane $Q_{s_0}$, $1\leq s_0\leq q_0$ that votes for both points ${\mathbf{a}}'_i$ and
${\mathbf{b}}_0$. It is easy to see that the broken line $\delta_{s_0}$ is the required one. Furthermore, on the
one hand, by Lemma 5.2, $Q\neq \bar{P}_i$ if and only if $Q$ votes against the point ${\mathbf{a}}'_i$. On the
other hand, the half-plane $Q$ votes against the endpoint ${\mathbf{a}}''_i\in{\cal{A}}$, ${\mathbf{a}}''_i\neq
{\mathbf{a}}'_i$, of the broken line $\delta_{s_0}$. Consequently, $Q\neq\bar{P}_i$ if and only if $Q$ votes
against the points ${\mathbf{a}}'_i$ and ${\mathbf{a}}''_i$.

This, in particular, implies that, if a half-plane of the collection $K'_1$ votes for two points ${\mathbf{a}}_i$
and ${\mathbf{b}}_i$, and also for two points ${\mathbf{a}}_j$ и ${\mathbf{b}}_j$, $i,j\in I_2$, $i\neq j$, then
this half-plane coincides with neither $\bar{P}_i$ nor $\bar{P}_j$. Denote by $p_1$ the number of such half-planes
contained in the collection $K'_1$. Consider also all such half-planes of this collection that vote for a certain
pair (and only for this pair) of points ${\mathbf{a}}_i$ or ${\mathbf{b}}_i$, $i\in I_2$, and denote by $p_2$ the
number of all these half-planes. Since $|I_2|=p-|I_1|=p-2q_0$, it follows that $2p_1+p_2\geq p-2q_0$. Let $p_3$ be
the number of members of the collection $K'_1$ present in $\bar{K}^{\ast}_0$. We use the following fact
(\cite{kobylkin2}): if a committee of a system of points ${\cal{D}}^{\ast}$ has $t$ members voting against some
two points of the system ${\cal{A}}$, then this committee contains $t+1$ members voting for both these points.
Consequently, $p_3\geq p_1$. Then, $q_1\geq p_1+p_2+p_3+q_0\geq 2p_1+p_2+q_0\geq p-q_0.$ The theorem is proved.

\end{document}